\documentclass[runningheads]{llncs}
\usepackage{amsmath}

\usepackage{amssymb}

\usepackage{etoolbox}

\usepackage{graphicx}

\usepackage[bookmarks,unicode,colorlinks=true]{hyperref}

\usepackage[capitalize]{cleveref}
\makeatletter
   \def\@citecolor{blue}%
   \def\@urlcolor{blue}%
   \def\@linkcolor{blue}%

\def\orcidID#1{\smash{\href{http://orcid.org/#1}{\protect\raisebox{-1.25pt}{\protect\includegraphics{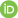}}}}}
\makeatother

\usepackage{cleveref}

\usepackage{listings}

\usepackage{mathpartir}

\usepackage{mathtools}

\usepackage{stmaryrd}

\usepackage{thmtools, thm-restate}

\usepackage{tikz}
\usetikzlibrary{decorations.pathreplacing,arrows.meta}

\usepackage[colorinlistoftodos]{todonotes}

\usepackage{underscore}

\usepackage{wasysym}

\usepackage{xargs}

\usepackage{xcolor}

\usepackage{xspace}

\newcommandx{\verified}[2][1=]{\todo[linecolor={\noteColor{verified}},backgroundcolor=cyan,bordercolor=black,size=\footnotesize,#1]{Verified: #2}}

\newcommandx{\corrected}[2][1=]{\todo[linecolor=green,backgroundcolor=green,bordercolor=black,size=\footnotesize,#1]{Fixed: #2}}

\newcommandx{\improve}[2][1=]{\todo[linecolor=yellow,backgroundcolor=yellow,bordercolor=black,size=\footnotesize,#1]{Improve: #2}}

\newcommandx{\change}[2][1=]{\todo[linecolor=red,backgroundcolor=red,bordercolor=black,size=\footnotesize,#1]{Change: #2}}

\newcommandx{\questionM}[2][1=]{\todo[linecolor=pink,backgroundcolor=pink,bordercolor=black,size=\footnotesize,#1]{Question: #2}}

\newcommandx{\suggestion}[2][1=]{\todo[linecolor=yellow,backgroundcolor=yellow,bordercolor=black,size=\footnotesize,#1]{Suggestion: #2}}

\usepackage{./packages/prooftree}

\newcommand{\keyword}[1]{\ensuremath{\mathsf{#1}}\xspace}
\newcommand{\letk}{\keyword{let}}
\newcommand{\boxk}{\keyword{box}}

\newcommand{\grammarEq}{\;::=\;\;}
\newcommand{\grammarOr}{\mid}

\newcommand{\emptyseq}{\varepsilon}

\newcommand{\sendOp}{{!}}
\newcommand{\receiveOp}{{?}}

\newcommand{\internalChoiceOp}{{\ensuremath{\oplus}}}
\newcommand{\externalChoiceOp}{{\ensuremath{\&}}}

\newcommand{\multOp}{m} 
\newcommand{\multun}{\omega}
\newcommand{\multlin}{\keyword1}

\newcommand{\Unit}{{\keyword{Unit}}}

\newcommand{\Wait}{{\keyword{Wait}}}
\newcommand{\Close}{{\keyword{Close}}}

\newcommand{\modeType}[3]{{\ensuremath{#1 #2 . #3}}}
\newcommand{\choiceType}[3]{{\ensuremath{#1 \{#2\}^{#3}}}}
\newcommand{\recursiveType}[2]{{\ensuremath{\mu #1 . #2}}}

\newcommand{\functionType}[3]{{\ensuremath{#1 \rightarrow_{#2} #3}}}
\newcommand{\productType}[2]{{\ensuremath{#1 \times #2}}}

\newcommand{\contextualType}[3]{\ensuremath{#1\, \vdash^{#2} #3}}
\newcommand{\contextualModalType}[1]{\square#1}

\newcommand{\binding}[2]{#1 \colon #2}
\newcommand{\csplit}{\circ} 
\newcommand{\cdiff}{\div}

\newcommand{\dualRelation}[2]{{#1\ \bot\ #2}}

\newcommand{\ctxlocal}[2]{#1^{<#2}}
\newcommand{\ctxouter}[2]{#1^{\ge #2}}
\newcommand{\ctxun}[1]{#1^\multun}
\newcommand{\ctxlin}[1]{#1^\multlin}
\newcommand{\ctxsplit}[2]{#1 \csplit #2}
\newcommand{\ctxdiv}[3]{#1 \stackrel{#2}{\leadsto} #3}
\newcommand{\ctxdiff}[2]{#1 \cdiff #2}
\newcommand{\ctxgetun}[1]{\mathcal{U}(#1)}
\newcommand{\ctxgetlin}[1]{\mathcal{L}(#1)}
\newcommand{\un}[1]{#1\:\multun}
\newcommand{\lin}[1]{#1\:\multlin}
\newcommand{\processCong}[2]{{\ensuremath{#1\ \equiv\ #2}}}
\newcommand{\ctype}[2]{#1\,\operatorname{ctype}^{#2}}

\newcommand{\fv}[1]{\ensuremath{\operatorname{free}#1}}
\newcommand{\dom}[1]{\ensuremath{\operatorname{dom}(#1)}}

\newcommand{\erase}[1]{\operatorname{erase}(#1)}

\newcommand{\constant}{c}
\newcommand{\evaluationContextVar}{E}

\newcommand{\unit}{\keyword{unit}}

\newcommand{\wait}{\keyword{wait}}
\newcommand{\close}{\keyword{close}}
\newcommand{\receive}{\keyword{receive}}
\newcommand{\send}{\keyword{send}}
\newcommand{\select}[1]{\keyword{select}\,#1}
\newcommand{\fork}{\keyword{fork}}
\newcommand{\fix}{\keyword{fix}}

\newcommand{\abstraction}[2]{{\ensuremath{\lambda #1 . #2}}}
\newcommand{\absT}[4]{{\ensuremath{\lambda_#1 #2 : #3 . #4}}}
\newcommand{\application}[2]{{\ensuremath{#1\,#2}}}
\newcommand{\pair}[2]{{\ensuremath{(#1, #2)}}}
\newcommand{\pairDestructor}[4]{{\ensuremath{\letk\ (#1, #2) = #3\ \keyword{in}\ #4}}}

\newcommand{\new}{\keyword{new}}
\newcommand{\matchk}{\keyword{match}}
\newcommand{\match}[3]{{\matchk\, #1\, \keyword{with}\, \{#2\}^{#3}}}

\newcommand{\cbox}[1]{\ensuremath{\boxk\,#1}} 
\newcommand{\letbox}[3]{\ensuremath{\letk~\boxk~#1=#2~\keyword{in}~#3}}

\newcommand{\appliedvar}[2]{#1[#2]}
\newcommand{\ctxval}[2]{#1.#2}
\newcommand{\etactx}[2]{\eta(#1, #2)}

\newcommand{\thread}[1]{{\ensuremath{\langle #1 \rangle}}}
\newcommand{\PAR}{\mid}
\newcommand{\parallelComposition}[2]{{\ensuremath{#1 \PAR #2}}}
\newcommand{\channelBinding}[3]{{\ensuremath{(\nu #1 #2) #3}}}

\newcommand{\hole}{[]}
\newcommand{\evaluationContextApply}[2]{#1 [#2]}

\newcommand{\typeof}[1]{\operatorname{typeof}(#1)}
\newcommand{\isterm}[3]{\ensuremath{#1 \vdash #2 : #3}}
\newcommand{\typesynth}[4]{\ensuremath{#1 \vdash #2 \Rightarrow #3 \mid #4}}
\newcommand{\isprocess}[2]{\ensuremath{#1 \vdash #2}}

\newcommand{\substitution}[2]{\{#1/#2\}} 

\newcommand{\rulename}[1]{\textsc{\small #1}\xspace}

\newcommand{\ruletsconst}{\rulename{T-Const}}
\newcommand{\ruletsvar}{\rulename{T-Var}}
\newcommand{\ruletsctxt}{\rulename{T-Ctx}}

\newcommand{\ruletsarrowuni}{\rulename{T-UnFunI}}
\newcommand{\ruletsarrowlini}{\rulename{T-LinFunI}}
\newcommand{\ruletsarrowe}{\rulename{T-FunE}}

\newcommand{\ruletspairi}{\rulename{T-PairI}}
\newcommand{\ruletspaire}{\rulename{T-PairE}}

\newcommand{\ruletsmatch}{\rulename{T-Match}}

\newcommand{\ruletsboxi}{\rulename{T-BoxI}}
\newcommand{\ruletsboxe}{\rulename{T-BoxE}}

\newcommand{\rulesubs}{\rulename{subs}}
\newcommand{\ruleweak}{\rulename{weak}}

\newcommand{\ruleatconst}{\rulename{A-Const}}
\newcommand{\ruleatvarun}{\rulename{A-UnVar}}
\newcommand{\ruleatvarlin}{\rulename{A-LinVar}}
\newcommand{\ruleatarrowuni}{\rulename{A-UnFunI}}
\newcommand{\ruleatarrowlini}{\rulename{A-LinFunI}}
\newcommand{\ruleatarrowe}{\rulename{A-FunE}}
\newcommand{\ruleatpairi}{\rulename{A-PairI}}
\newcommand{\ruleatpaire}{\rulename{A-PairE}}
\newcommand{\ruleatmatch}{\rulename{A-Match}}
\newcommand{\ruleatboxi}{\rulename{A-BoxI}}
\newcommand{\ruleatboxe}{\rulename{A-BoxE}}
\newcommand{\ruleatctxt}{\rulename{A-Ctx}}

\newcommand{\rulepfexp}{\rulename{R-Exp}}
\newcommand{\rulepfpar}{\rulename{P-Par}}
\newcommand{\rulepfbind}{\rulename{P-Res}}

\newcommand{\ruletrbeta}{\rulename{R-Beta}}
\newcommand{\ruletrsplit}{\rulename{R-Split}}
\newcommand{\ruletrctx}{\rulename{R-Ctx}}
\newcommand{\ruletrletbox}{\rulename{R-LetBox}}
\newcommand{\ruletrfix}{\rulename{R-Fix}}

\newcommand{\ruleprfork}{\rulename{R-Fork}}
\newcommand{\ruleprexp}{\rulename{R-Exp}}
\newcommand{\ruleprnew}{\rulename{R-New}}
\newcommand{\ruleprcom}{\rulename{R-Com}}
\newcommand{\ruleprchoice}{\rulename{R-Branch}}
\newcommand{\ruleprclose}{\rulename{R-Close}}
\newcommand{\ruleprpar}{\rulename{R-Par}}
\newcommand{\ruleprbind}{\rulename{R-Res}}
\newcommand{\ruleprcong}{\rulename{R-Struct}}

\newcommand{\redOp}{\rightarrow}
\newcommand{\reduction}[2]{\ensuremath{#1 \redOp #2}}
\newcommand{\processReduction}[2]{\ensuremath{#1 \redOp #2}}

\newcommand{\ie}{i.e.,\xspace} 
\newcommand{\eg}{e.g.,\xspace}  

\newcommand{\etal}{et al.\xspace} 

\definecolor{darkviolet}{rgb}{0.5,0,0.4}
\definecolor{darkgreen}{rgb}{0,0.4,0.2}
\definecolor{darkblue}{rgb}{0.1,0.1,0.9}
\definecolor{darkgrey}{rgb}{0.5,0.5,0.5}
\definecolor{lightblue}{rgb}{0.4,0.4,1}

\lstdefinestyle{eclipse}{
  breaklines=true,
  basicstyle=\sffamily\small,
  emphstyle=\color{red}\bfseries,
  keywordstyle=\color{darkviolet}\bfseries,
  commentstyle=\color{darkgreen},
  stringstyle=\color{darkblue},
  numberstyle=\color{darkgrey},
  emphstyle=\color{red},
  showstringspaces=false,
}

\tikzset{
arrow/.style={>={Stealth[inset=0pt,length=8pt,angle'=28,round]}}}

\usepackage{prooftree}

\usepackage{iftex}

\begin{document}
\title{Contextual Metaprogramming for Session Types}
%
%
\author{Pedro Ângelo\inst{1}\orcidID{0000-0002-7849-195X} \and
Atsushi Igarashi\inst{2}\orcidID{0000-0002-5143-9764} \and
Yuito Murase\inst{2}\orcidID{0000-0001-6038-6249} \and
Vasco T. Vasconcelos\inst{1}\orcidID{0000-0002-9539-8861}}
%
%
\institute{LASIGE, Faculdade de Ciências da Universidade de Lisboa, Portugal\\
\email{\{pjangelo,vmvasconcelos\}@ciencias.ulisboa.pt}%
\and
Kyoto University, Kyoto, Japan\\
\email{\{murase@fos.,igarashi@\}kuis.kyoto-u.ac.jp}}

\maketitle

\begin{abstract}
We propose the integration of staged metaprogramming into a session-typed message passing functional language.
We build on a model of contextual modal type theory with multi-level contexts, where contextual values, closing arbitrary terms over a series of variables, may be boxed and transmitted in messages.
Once received, one such value may then be unboxed and locally applied before being run.
To motivate this integration, we present examples of real-world use cases, for which our system would be suitable, such as servers preparing and shipping code on demand via session typed messages.
We present a type system that distinguishes linear (used exactly once) from unrestricted (used an unbounded number of times) resources, and further define a type checker, suitable for a concrete implementation.
We show type preservation, a progress result for sequential computations and absence of runtime errors for the concurrent runtime environment, as well as the correctness of the type checker.
\end{abstract}


\section{Introduction}
\label{sec:intro}
Metaprogramming manipulates code in order to generate and evaluate code at
runtime, allowing compilers in particular to explore the availability of certain arguments
to functions in order to save computational effort.
%
In this paper we are interested in programming languages where the code produced
is typed by construction and where code may refer to a context providing types
for the free variables, commonly known as contextual
typing~\cite{DBLP:journals/pacmpl/JangGMP22,DBLP:conf/esop/MuraseNI23,DBLP:journals/jfp/NanevskiP05,DBLP:journals/tocl/NanevskiPP08}.
On an orthogonal axis, session types have been advocated as a means to discipline
concurrent computations, by accurately describing protocols for the channels
used to exchange messages between processes~\cite{DBLP:journals/iandc/AlmeidaMTV22,DBLP:journals/jfp/GayV10,DBLP:conf/concur/Honda93,DBLP:conf/esop/HondaVK98,DBLP:conf/parle/TakeuchiHK94,DBLP:conf/icfp/ThiemannV16,DBLP:journals/iandc/Vasconcelos12}.

The integration of session types with metaprogramming allows one to set up
code-producing servers that run in parallel with the rest of the program and
provide code on demand, exchanged via typed channels.
Linearity is central to session types, but current metaprogramming models lack
support for such a feature. We extend a simple model of contextual modal type
theory with support for session types, to obtain a call-by-value linear lambda
calculus with multi-level contexts.

We briefly discuss the design decisions we have taken for, when combining
session types and metaprogramming, we were faced with several possible
theoretical foundations for each component, each with its own advantages and
disadvantages.

We base our development on Davies and
Pfenning~\cite{DBLP:journals/jacm/DaviesP01} where a box modality distinguishes
generated code. We further allow code to refer to variables in a context,
described by contextual types, along the lines of Nanevski
\etal~\cite{DBLP:journals/tocl/NanevskiPP08}. The M\oe bius language, proposed
by Jang \etal~\cite{DBLP:journals/pacmpl/JangGMP22}, further adds to modal
contextual type theory the provision for pattern matching on code, for
generating polymorphic code, and for generating code that depends on other code
fragments. We forgo the first two directions, and base our development on the
last.
We propose a multilevel contextual modal linear lambda calculus with support for
session types, where in particular the composition of code fragments avoids
creating extraneous administrative redexes due to boxing and unboxing.

An alternative starting point would have been the Fitch- or Kripke-style
formulation, providing for the Lisp quote/unquote, where typing contexts are
viewed as stacks modeling the different stages of
computation~\cite{DBLP:conf/fossacs/Clouston18,DBLP:conf/esop/MuraseNI23,DBLP:journals/pacmpl/ValliappanRC22}.
It seemed to us that the let-box approach would simplify the extension to the linear setting and to session
types.
Or, temporal types~\cite{DBLP:journals/jacm/Davies17} could be another alternative approach, which is adoped by MetaML and many staged programming languages~\cite{DBLP:journals/corr/abs-2309-08207,DBLP:conf/gpce/StuckiBO21,DBLP:conf/popl/TahaN03,DBLP:journals/pacmpl/XieWNY23}. Temporal type systems, however, are known to be unsound in the presence of computational effects~\cite{DBLP:journals/jfp/KameyamaKS11,DBLP:conf/esop/Rhiger12,DBLP:conf/popl/TahaN03}, whereas channel-based communication is inherently effectful. We therefore preferred contextual modal type–based systems such as M\oe bius, which are compatible with session types. 

As for session types, our formulation largely follows that of Gay and
Vasconcelos~\cite{DBLP:journals/jfp/GayV10} that elegantly integrates functional
programming with a process calculus for the runtime, providing an adequate
formal model for message passing concurrent programming languages.
Programs are written in a conventional, intuitive, programming language that
does not force programmers to resort to the level of processes, as is the case
of a recent proposal by Sano \etal~\cite{DBLP/conf/icfp/SanoGKBT2025}.
An alternative starting point would have been a linear logic inspired language,
providing for deadlock freedom, as proposed by Caires, Pfenning, Wadler~\cite{DBLP:conf/concur/CairesP10,DBLP:journals/jfp/Wadler14}.
However, enforcing the absence of deadlocks along the ideas of the
aforementioned works would impose significant syntactic restrictions, which we
prefer to avoid in favour of a more flexible design.

It is worth emphasizing that our contribution is not merely to combine M\oe bius
and the language of Gay and Vasconcelos. In order to integrate the two, we
needed to introduce (message passing) concurrency and a linear type system
capable of handling both (recursive) linear and unrestricted resources. This
extension forms one of the key technical contributions of this work.

The rest of the paper is structured as follows. In \cref{section:motivation} we
motivate our proposal by means of examples. \Cref{section:expressions-processes}
introduces terms, processes, term evaluation and process reduction.
\Cref{section:types} describes types and a type assignment system.
\Cref{section:algorithmic-type-checking} introduces type checking.
\Cref{sec:related} reviews related work. \Cref{section:conclusion} concludes the
paper and points directions for future work. We leave for the appendix extra
examples (\cref{sec:extra-examples}), and proofs for all results
(\cref{sec:proofs}).


\section{Motivation}
\label{section:motivation}

We begin by informally introducing the language proposed in this paper through several examples. First, we present a small program in which processes send and receive code fragments over channels and execute them, illustrating the basic ideas of the language’s syntax and type system. We then discuss how the language can model various real-world use cases that involve transmitting code fragments via channels.

We start by presenting the generation of a code fragment to send a fixed number
of integer values on a stream. The type of streams, as seen from the side of
processes writing on the stream, is as follows.
\begin{lstlisting}
type Stream = oplus{More: !Int.Stream, Done: Close}
\end{lstlisting}
The writer chooses between selecting \lstinline|More| values or selecting
\lstinline|Done|. In the former case, the writer sends an integer value
(\lstinline|!Int|) and ``goes back to the beginning'' (\lstinline|Stream|); in
the latter case the writer must close the channel (\lstinline|Close|).
Function \lstinline|sendFives| below accepts an integer value and returns a code
fragment. The code fragment requires a channel endpoint (of type
\lstinline|Stream|) and, when executed, produces a unit value; its type is
written \lstinline|[Stream vdash Unit]|, which is called a \emph{box type}. We
proceed by pattern-matching on the integer parameter.
\begin{lstlisting}
sendFives : Int -> [StreamvdashUnit]
sendFives 0 = box (y. close (select Done y))
sendFives n = let box u = sendFives (n - 1)
              in box (x. u[send 5 (select More x)])
\end{lstlisting}

When all values have been sent on the stream (when \lstinline|n| is
\lstinline|0|), all it remains is to
\lstinline|select Done| and then close the channel. The \lstinline|box| term
generates code under a variable environment (an evaluation context), in this
case containing variable \lstinline|y| alone, denoting the channel endpoint.
The code fragment \lstinline|y. close (select Done y)| enclosed inside \lstinline|box|
is called a \emph{contextual value}.

When there are values left to be sent (when \lstinline|n| is different from
\lstinline|0|), we recursively compute code to send \lstinline|n-1| values,
unbox it storing the resulting contextual value (say \lstinline|z.close (select Done z)|) in \lstinline|u|, and then prepare code to send the
\lstinline|n|-th value.  The use of the let-box bound variable \lstinline|u|, which is given
a \emph{contextual type} \lstinline|Stream vdash Unit| (without square brackets \lstinline|[ ]|),
always takes the form \lstinline|u[e]|, where \lstinline|e| is
a term of type \lstinline{Stream}, and results in a term
of type \lstinline|Unit| obtained by substituting \lstinline|e| (without being evaluated) for \lstinline|z|.
The part \lstinline|[e]| is called an \emph{explicit substitution}.
In this example, term \lstinline|u[send 5 (select More x)]| applies
\lstinline|u| of type \lstinline|Stream vdash Unit| to
term \lstinline|send 5 (select More x)| of type
\lstinline|Stream|.
If \lstinline|u| is the contextual value \lstinline|z.close (select Done z)|,
then \lstinline|box x.u[send 5 (select More x)]| evaluates to
\lstinline|box x. close (select Done (send 5 (select More x)))|, a piece of code that
sends number \lstinline|5| on channel \lstinline|x| and then closes the channel.

The main difference between a contextual type such as \lstinline|Stream vdash Unit|
and the corresponding box type (\lstinline|[Stream vdash Unit]|) is that
the latter kind of types represents \emph{first-class} code values---one can pass it to another function
or store into a data structure---whereas the former kind are second-class---a
variable of type
\lstinline|Stream vdash Unit| may only be used with an explicit substitution
to compose another piece of code.
See work on contextual modal type theory (\eg~\cite{DBLP:journals/tocl/NanevskiPP08}) for more details.


We may now compute and run code to send a fixed number of integer values.
\begin{lstlisting}
send4Fives : Stream -> Unit
send4Fives c = let box u = sendFives 4 in u[c]
\end{lstlisting}
%
Term \lstinline|sendFives 4| is a boxed code fragment (a term) of type
\lstinline|[Stream vdash Unit]|. Then, \lstinline|u| is an unboxed code fragment
(a contextual value) of contextual type \lstinline|Stream vdash Unit|. We
provide the contextual value with an explicit substitution \lstinline|[c]|. The
whole \lstinline|let| term then amounts to running the code
\begin{lstlisting}
  close (select Done (send 5 (select More (...send 5 (select More c)...))))
\end{lstlisting}
without calling function \lstinline|sendFives| or using recursion in any other form.







The next example transmits code on channels. Imagine a server preparing code on
behalf of clients. The server uses a channel to interact with its clients: it
first receives a number \lstinline|n|, then replies with code to send
\lstinline|n| fives, and finally waits for the channel to be closed.
The type of the communication channel is as follows.
\begin{lstlisting}
type Builder = ?Int.![StreamvdashUnit].Wait
\end{lstlisting}

The server receives \lstinline|n| on a given channel and computes the code using
a call to \lstinline|sendFives|. It then waits for the channel to be closed.
\begin{lstlisting}
serveFives : Builder -> Unit
serveFives c =
  let (n,c) = receive c in wait (send (sendFives n) c)
\end{lstlisting}

On the other end of the channel sits a client: it sends a number (4 in this
case), receives the code (of type \lstinline|[Stream vdash Unit]|), closes the
channel and evaluates the code received.
\begin{lstlisting}
sendFives' : Dual Builder -> Stream -> Unit
sendFives' c d =
  let (code,c) = receive (send 4 c) in close c ;
  let box u = code in u[d]
\end{lstlisting}
The \lstinline|Dual| operator on session types provides a view of the other end
of the channel. In this case, \lstinline|Dual Builder| is the type
\lstinline|!Int.?[Stream vdash Unit].Close|, where \lstinline|!| is turned into
\lstinline|?| and \lstinline|Wait| is turned into
\lstinline|Close| (and conversely in both cases).
Notice that \lstinline|code| is a boxed code fragment of type 
\lstinline|[Stream vdash Unit]|, hence \lstinline|u| is the corresponding code
fragment (of type \lstinline|Stream vdash Unit|) and \lstinline|u[d]| runs the
code on channel \lstinline|d|.

To complete the example we need a function for reading streams, that is, a
consumer of type \lstinline|Dual Stream -> Unit|. Function \lstinline|readInts|
reads and discards all integer values on the stream and then waits for the
stream to be closed. Here we proceed by pattern matching on the label received
on the channel.
\begin{lstlisting}
readInts : Dual Stream -> Unit
readInts (Done c) = wait c
readInts (More c) = let (_, c) = receive c in readInts c
\end{lstlisting}

Finally, the main thread forks two threads---one running \lstinline|serveFives|,
the other to collect the integer values (\lstinline|readInts|)---and continues with
\lstinline|sendFives'|. We take advantage of a primitive function,
\lstinline|forkWith| that expects a suspended computation (a thunk), creates a
new channel, forks the thunk on one end of the channel, and returns the other
end of the channel for further interaction.
\begin{lstlisting}
main : Unit
main =
  let c = forkWith (lambda_.serveFives) in -- c : Dual Builder
  let d = forkWith (lambda_.readInts) in   -- d : Stream
  sendFives' c d
\end{lstlisting}

\begin{figure}[t!]
  \begin{center}
    \begin{tikzpicture}
      \node (process1Begin) [align=center] {\lstinline|serveFives|}; \node
      (process1End) [below of=process1Begin, node distance=3.85cm] {}; \draw
      [draw, very thick, -] (process1Begin.south) -- (process1End.north);

      \node (process2Begin) [align=center, right of=process1Begin, node
      distance=4cm] {\lstinline|sendFives'|}; \node (process2End) [below
      of=process2Begin, node distance=3.85cm] {}; \draw [draw, very thick, -]
      (process2Begin.south) -- (process2End.north);

      \node (process3Begin) [align=center, right of=process2Begin, node
      distance=4cm] {\lstinline|readInts|}; \node (process3End) [below
      of=process3Begin, node distance=3.85cm] {}; \draw [draw, very thick, -]
      (process3Begin.south) -- (process3End.north);

      \draw [decorate,decoration={brace,amplitude=10pt}] (process1Begin.north)
      -- (process2Begin.north) node[near start, xshift=-1cm,
      yshift=0.8cm]{\lstinline|Builder|}
      node[midway,yshift=0.8cm]{\lstinline|c|} node[near end, xshift=1cm,
      yshift=0.8cm]{\lstinline|Dual Builder|};

      \draw [decorate,decoration={brace,amplitude=10pt}] (process2Begin.north)
      -- (process3Begin.north) node[near start, xshift=-1cm,
      yshift=1.2cm]{\lstinline|Stream|} node[midway,yshift=1.2cm]{\lstinline|d|}
      node[near end, xshift=1cm, yshift=1.2cm]{\lstinline|Dual Stream|};

      \draw [draw, <-, arrow] ([yshift=-.25cm]process1Begin.south) -- node
      [above] {\lstinline|4|} ([yshift=-.25cm]process2Begin.south); \draw [draw,
      ->, arrow] ([yshift=-.75cm]process1Begin.south) -- node [above]
      {\lstinline|code|} ([yshift=-.75cm]process2Begin.south); \draw [draw, <-,
      arrow] ([yshift=-1.25cm]process1Begin.south) -- node [above]
      {\lstinline|close|} ([yshift=-1.25cm]process2Begin.south);

      \draw [draw, ->, arrow] ([yshift=-1cm]process2Begin.south) -- node [above]
      {\lstinline|More|} ([yshift=-1cm]process3Begin.south); \draw [draw, ->,
      arrow] ([yshift=-1.5cm]process2Begin.south) -- node [above]
      {\lstinline|5|} ([yshift=-1.5cm]process3Begin.south); \draw [draw, ->,
      arrow] ([yshift=-2cm]process2Begin.south) -- node [above]
      {\lstinline|More|} ([yshift=-2cm]process3Begin.south); \draw [draw, ->,
      arrow] ([yshift=-2.5cm]process2Begin.south) -- node [above] {\dots}
      ([yshift=-2.5cm]process3Begin.south); \draw [draw, ->, arrow]
      ([yshift=-3cm]process2Begin.south) -- node [above] {\lstinline|Done|}
      ([yshift=-3cm]process3Begin.south); \draw [draw, ->, arrow]
      ([yshift=-3.5cm]process2Begin.south) -- node [above] {\lstinline{close}}
      ([yshift=-3.5cm]process3Begin.south);
    \end{tikzpicture}
  \end{center}
  \caption{Message sequence chart for the \lstinline|SendFives| example}
  \label{fig:msc}
\end{figure}
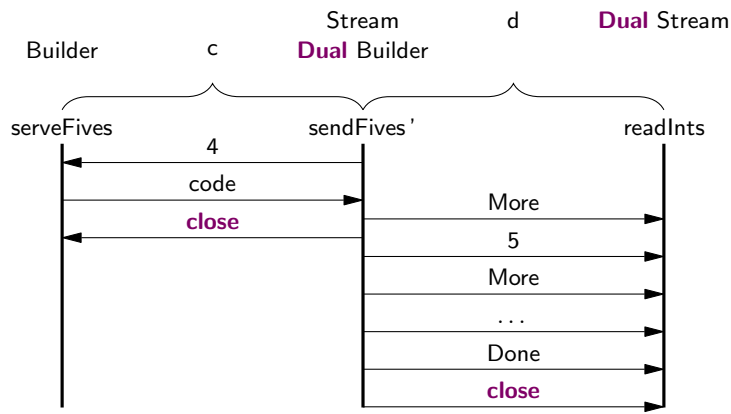

The interaction among the three processes is depicted in \cref{fig:msc}, where
the \lstinline|code| fragment produced by function \lstinline|serveFives| and
transmitted to \lstinline|sendFives'| is
\begin{lstlisting}
  box (c.close (select Done (send 5 (select More (...(select More c)...)))))
\end{lstlisting}

Contextual metaprogramming for session types runs on the top of a linear type
system: each resource is classified as \emph{linear} (used exactly once) or
\emph{unrestricted} (used an unbounded number of times, including zero). Channel
endpoints are always linear so that protocols may not encounter unexpected
interactions. Functions may be linear or unrestricted. All the functions we have
seen so far are unrestricted because they do not capture free linear values. The
recursive functions (\lstinline|sendFives| for example) are necessarily
unrestricted, for they are discarded at the end of recursion. Some other values
could be classified as linear if so desired. One such example is function
\lstinline|serveFives| which is used exactly once in function \lstinline|main|.
We annotate arrows with $\multun$ for unrestricted and $\multlin$ for linear. In
examples, we often omit the $\multun$ label. Function \lstinline|serveFives|
could as well be of type \lstinline|Builder arrowlin Unit|.

Boxed code fragments are always unrestricted, so that they may be used as many
times as needed.
Below are examples where a
code fragment (denoted by \lstinline|u| and of type
\lstinline|Stream vdash Stream|) that sends \lstinline|5| on a
given channel is duplicated (first example) or discarded (second example).
\begin{lstlisting}
sendTwice : Stream -> Unit
sendTwice = let box u = box (y. send 5 (select More y))
            in lambdax.close (select Done (u[u[x]])
sendNone : Stream -> Unit
sendNone = let box u = box (y. send 5 (select More y))
           in lambdax.close (select Done x)
\end{lstlisting}
In the first case, two \lstinline|5| messages are sent on channel \lstinline|x|
and then the channel is closed; in the second, the only interaction on the
channel is to close it.
\Cref{sec:extra-examples} contains an example of  multiple level programming.

The \lstinline|sendFives| example is somewhat artificial and serves mainly to
illustrate the basic idea of combining session types with metaprogramming.
Nevertheless, this combination naturally appears in various forms in real-world
programs, which we now discuss.


\textbf{Distributed computing.}
In distributed systems, the computational power of several networked machines, which need not be in the same geographical location or network, is harnessed towards the completion of a shared goal.
To distribute tasks between nodes, messages, containing code to be executed, are
dispatched across the network. Commercial implementations such as Apache Spark
\cite{DBLP:conf/hotcloud/ZahariaCFSS10} and Hadoop
\cite{DBLP:books/daglib/0035689}, follow the \emph{MapReduce} programming model
\cite{DBLP:journals/cacm/DeanG08}, inspired by functional programming. A related
concept is that of \emph{volunteer computing}, where volunteers (\eg personal
laptops) donate their idle processing power to solve large-scale problems.
Boinc~\cite{boinc}, targeting scientific projects with intensive computational
needs, is one such example: volunteers subscribe to receive tasks as code,
compute it and then return the result.
\par
Consider the following protocol, structuring the communication between the
central server and a volunteer, in the context of a prime search computational
effort, such as PrimeGrid~\cite{primegrid}.
\begin{lstlisting}
type PrimeTask = &{ Task: ![Int vdash Bool].PrimeTask,
                    Test: !Int.?Bool.PrimeTask,
                    Done: Wait }
\end{lstlisting}
The server offers three alternatives to subscribed volunteers.
The first alternative is \lstinline|Task|, which will provide the volunteer with a task to be performed.
The task itself is specified as the box type \lstinline|[Int vdash Bool]|.
The task is then sent to the volunteer (\lstinline|![Int vdash Bool]|), and then the protocol continues from the beginning (\lstinline|PrimeTask|).
Afterwards, the volunteer can donate its idle computational resources by selecting \lstinline|Test|.
The server sends a number (\lstinline|!Int|), which we will call the \emph{prime candidate}, to be tested,
and then expects a result, \ie either \lstinline|True| or \lstinline|False| in return (\lstinline|?Bool|), which the volunteer stores.
The interaction again goes back to the beginning.
The server provides a third alternative, \lstinline|Done|, allowing the volunteer to stop contributing, subsequently waiting for the volunteer to close the connection (\lstinline|Wait|).

Assuming a \lstinline|testPrime| function of type \lstinline|[Int vdash Bool]|, \ie code that accepts the splicing in of a number and returns either \lstinline|True| or \lstinline|False|, we write the server as follows:
\begin{lstlisting}
server : [Int] -> [Bool] -> PrimeTask -> [Int]
server cs res (Task s) =
  let s = send testPrime in server cs res s
server (c:cs) res (Test s) =
  let s = send c in
  let (result, s) = receive s in
  server cs (res ++ [result]) s
server _ res (Done s) = wait s ; res
\end{lstlisting}

A volunteer, when connected to the server, selects between these three options.
By selecting \lstinline|Task| and then received the code, the volunteer stores it for later use.
By selecting \lstinline|Test|, the volunteer receives the prime candidate and runs the test by splicing the candidate into the code.
The result is then sent to the server.
The volunteer may repeat this action as many times as he wishes, thus donating more processing time.
Finally, when the volunteer no longer wishes to contribute, he can just select \lstinline|Done| and close the connection.
\begin{lstlisting}
volunteer : Dual PrimeTask -> Unit
volunteer v  = 
  let v = select Task v in
  let (primeTest, v) = receive v in
  let v = select Test v in
  let (cand, v) = receive v in
  let box t = primeTest in send (t[cand]) v in
  let v = select Test v in
  let (cand, v) = receive v in
  let box t = primeTest in send (t[cand]) v in
  ...
  let v = select Done v in close v
\end{lstlisting}

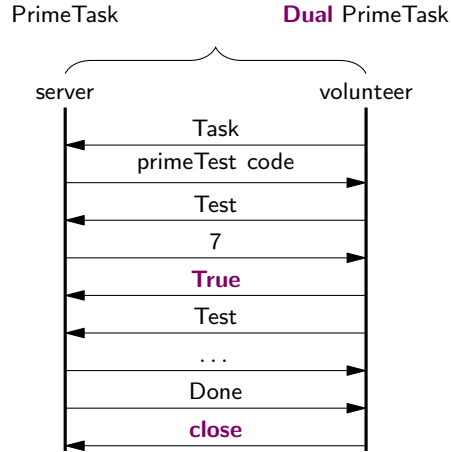
\begin{figure}[t!]
\begin{center}
\begin{tikzpicture}
\node (process1Begin) [align=center] {\vphantom{l}\lstinline|server|};
\node (process1End) [below of=process1Begin, node distance=4.91cm] {};
\draw [draw, very thick, -] (process1Begin.south) -- (process1End.north);

\node (process2Begin) [align=center, right of=process1Begin, node distance=4cm] {\lstinline|volunteer|};
\node (process2End) [below of=process2Begin, node distance=4.91cm] {};
\draw [draw, very thick, -] (process2Begin.south) -- (process2End.north);

\draw [decorate,decoration={brace,amplitude=10pt}]  (process1Begin.north) -- (process2Begin.north) node[near start, xshift=-1cm, yshift=0.8cm]{\lstinline|PrimeTask|} node[midway,yshift=0.8cm]{
} node[near end, xshift=1cm, yshift=0.8cm]{\lstinline|Dual PrimeTask|};

\draw [draw, <-, arrow] ([yshift=-.5cm]process1Begin.south) -- node [above] {\lstinline|Task|} ([yshift=-.5cm]process2Begin.south);
\draw [draw, ->, arrow] ([yshift=-1cm]process1Begin.south) -- node [above] {\lstinline|primeTest\ code|} ([yshift=-1cm]process2Begin.south);
\draw [draw, <-, arrow] ([yshift=-1.5cm]process1Begin.south) -- node [above] {\lstinline|Test|} ([yshift=-1.5cm]process2Begin.south);
\draw [draw, ->, arrow] ([yshift=-2cm]process1Begin.south) -- node [above] {\lstinline|7|} ([yshift=-2cm]process2Begin.south);
\draw [draw, <-, arrow] ([yshift=-2.5cm]process1Begin.south) -- node [above] {\lstinline|True|} ([yshift=-2.5cm]process2Begin.south);
\draw [draw, <-, arrow] ([yshift=-3cm]process1Begin.south) -- node [above] {\lstinline|Test|} ([yshift=-3cm]process2Begin.south);
\draw [draw, ->, arrow] ([yshift=-3.5cm]process1Begin.south) -- node [above] {\dots} ([yshift=-3.5cm]process2Begin.south);
\draw [draw, ->, arrow] ([yshift=-4cm]process1Begin.south) -- node [above] {\lstinline|Done|} ([yshift=-4cm]process2Begin.south);
\draw [draw, <-, arrow] ([yshift=-4.5cm]process1Begin.south) -- node [above] {\lstinline|close|} ([yshift=-4.5cm]process2Begin.south);
\end{tikzpicture}
\end{center}
\caption{Message sequence chart for the \lstinline|PrimeTask| example}
\label{fig:msc-primetask}
\end{figure}

When running our main program, the interaction unfolds as depicted in \cref{fig:msc-primetask}.
\begin{lstlisting}
main : [Bool]
main = let s = forkWith (lambda_.volunteer) in -- s : Dual PrimeTask
       server [7..] s
\end{lstlisting}

However, during the course of the interaction, the primality test algorithm can be improved.
Hence, the volunteer can regularly poll for updated code, by selecting \lstinline|Task| again.
\begin{lstlisting}
volunteer : Dual PrimeTask -> Unit
volunteer v  = 
  ...
  let v = select Task v in
  let (primeTest, v) = receive v in
  ...
\end{lstlisting}

In this example, 
session types are used to govern the communication aspect of task assignment, whereas staged metaprogramming models sending code via messages.


\textbf{Computation offloading}
consists in transferring computationally intensive tasks from
resource-constrained devices (drones, IoT devices, smartphones) to more powerful
remote servers. The main advantage lies in circumventing the shortcomings of
these devices, particularly low storage capacity, low processing power and low
battery life, enabling greater efficiency and decreased energy consumption on
the device, as witnessed by Clonecloud~\cite{DBLP:conf/eurosys/ChunIMNP11}. The
advantages of this concept become highly relevant in the context of high-throughput
networks, since the overhead costs regarding the communication aspect
are lessened. Our contribution relates closely with the concept of computation
offloading, where session types broker the offloading of tasks, described as
code via box types, as in the type below.
\begin{lstlisting}
type ComputeServer = &{ Compute:?[vdashInt].!Int.ComputeServer
                      , Done:Wait }
\end{lstlisting}

\textbf{Code generation service.} Template metaprogramming, as proposed by Sheard and Peyton Jones \cite{DBLP:conf/haskell/SheardJ02} for Haskell, allows one to write metaprograms that can produce other programs, such as those capable of manipulating arbitrary-sized data structures.
For example, in Haskell, it is possible to write metaprograms such as: \lstinline{fstN}, which produces functions to extract the first element of an arbitrary-sized tuple, or \lstinline{zipN}, which produces n-ary zip functions.
Consider the protocol for a server offering code generating services:
\begin{lstlisting}
type CodeGenerator = &{
  Fst3: ![vdash (a,b,c) -> a].CodeGenerator,
  Fst4: ![vdash (a,b,c,d) -> a].CodeGenerator,
  ...
  Zip3: ![vdash [a] -> [b] -> [c] -> [(a,b,c)]].CodeGenerator
  ...
  Done: Close}
\end{lstlisting}
The server provides clients with functions to manipulate data structures of arbitrary size.
To satisfy a request, the server evaluates the specified template metaprogram, splicing in the size of the data structure, and sending the result to the client.
By only storing the function templates, and deriving the instances, \ie by running the metaprograms, the server's storage requirements remain low as the service is scaled with more function instances it offers.
While this example is still not feasible, it serves to illustrate the advantages of our contribution.


\textbf{Other use cases.}
Multi-tier programming languages, such as Hop.js \cite{DBLP:conf/icfp/SerranoP16}, Ocsigen/Eliom \cite{DBLP:conf/ml/Balat06} , Meteor.js \cite{meteor.js} and Unison \cite{unison}, allow a single program to describe the functionality of the different tiers of an application, \eg client, server and database tier, in one unifying language.
Session types allow to structure and verify client-server interactions, while providing type-safety properties compatible with multi-tier programming.
ScalaLoci \cite{scalaloci} and Ur/Web \cite{urweb} are languages which relate to multi-staged programming, but lack explicit communication models.
These would benefit from session types, \eg to ensure correctness across computation stages.

HTMX~\cite{htmx}, Next.js~\cite{nextjs} and React Server Components~\cite{rsc}
are web frameworks featuring pre-rendering strategies, where the server
pre-renders parts of the UI and sends them to the client, while also allowing
server components, \ie code that only runs in servers. Session types allow to
verify the correctness of UI update flows, while staged metaprogramming helps
establish the theoretical basis behind the pre-rendering strategies.

\section{Terms, processes  and operational semantics}
\label{section:expressions-processes}

We build our language on the syntactic categories of variables, $x,y,z$ and of
labels $l$.
We write $\overline X$ for a sequence of objects $X_1\cdots X_n$ with $n\ge0$.
The empty sequence (when $n=0$) is denoted by~$\emptyseq$.
%
%
The syntax of terms is in \cref{figure:expressions-processes}.
\begin{figure*}[t!]
  \begin{alignat*}{4}
    &\text{Constant} &&\constant  && \grammarEq  &&%
    \close \grammarOr
    \wait \grammarOr
    \send \grammarOr
    \receive \grammarOr
    \select{l} \grammarOr
    \new \grammarOr
    \fork \grammarOr
    \fix \grammarOr
    \unit
    \\
    &\text{Contextual value}  &&\rho,\sigma  && \grammarEq  &&%
    \ctxval{\overline x}M
    \\
    &\text{Value}  &&v,u  && \grammarEq  &&
    c \grammarOr
    \appliedvar x{\emptyseq} \grammarOr
    \abstraction{x}{M} \grammarOr
    \pair{v}{v} \grammarOr
    \cbox \sigma \grammarOr
    \application{\send}{v}
    \\
    &\text{Term}  &&M, N && \grammarEq  &&
    v \grammarOr
    \appliedvar x{\overline \sigma} \grammarOr
    \application{M}{M} \grammarOr
    \pair{M}{M}  \grammarOr
    \pairDestructor{x}{x}{M}{M} \grammarOr
    \\ &&&&&&&
    \letbox xMM \grammarOr
    \match{M}{l \rightarrow M_l}{l \in L}
    \\
    &\text{Process}  &&P, Q  && \grammarEq  &&
    \thread{M} \grammarOr
    \parallelComposition{P}{Q} \grammarOr
    \channelBinding{x}{x}{P}
    \\
    &\text{Evaluation context }  &&E  && \grammarEq  &&
    \hole \grammarOr%
    \application{E}{M} \grammarOr%
    \application{v}{E} \grammarOr%
    \pair{E}{M} \grammarOr%
    \pair{v}{E} \grammarOr%
    \pairDestructor{x}{x}{E}{M} \grammarOr%
    \\ &&&&&&&
    \match{E}{l \rightarrow M_l}{l \in L} \grammarOr \letbox {x}{E}{M}
  \end{alignat*}
  \caption{Terms and processes}
  \label{figure:expressions-processes}
\end{figure*}
%
%
It includes channel related primitives: \close for closing a channel, \wait for waiting for
a channel to be closed, \send to write a given value on a channel, \receive to
read a value from a channel, $\select l$ for sending label $l$ on a given
channel and \new for creating a channel.
\fork creates a new thread, \fix is the call-by-value fixed point combinator.
\unit is the only value of its type, and can be thought as a placeholder for other primitive values, such as integer or boolean values found in examples.

Further terms include constants, modal variables (playing the dual role of term
variables and of channel endpoints), lambda abstraction and
application, pair introduction and elimination, box introduction and
elimination, a \matchk term to branch according to the label $l$ selected.
The examples in \cref{section:motivation} use \lstinline|forkWith|, a convenient
function that puts together channel creation and process spawning:
\begin{lstlisting}
  forkWith f = let (x,y) = new () in fork (lambda_.f y) ; x
\end{lstlisting}

To objects of the form $\ctxval{\overline x}M$ we call \emph{contextual values}.
They denote code fragments $M$ parameterized by the variables in sequence
$\overline x$.
A \emph{contextual term variable} $\appliedvar{x}{\overline\sigma}$ applies the
code fragment described by $x$ to contextual values $\overline{\sigma}$.
%
The $\cbox \sigma$ term turns a code fragment denoted by a contextual value
$\sigma$ into a term; the $\letbox xMN$ term eliminates the box in $M$ and binds
$x$ to it, to be used in $N$.
The \letk-\boxk term allows code to be spliced into another code fragment ($N$), by
eliminating the box in $M$ and binding the result.
When $\overline x$ is the empty sequence we sometimes write $M$ in
place of the contextual value $\ctxval \emptyseq M$. Similarly, when
$\overline\sigma$ is the empty sequence we sometimes write $x$ instead
of the contextual term variable $\appliedvar{x}{\emptyseq}$.  Channel
endpoints are always expressed as terms $\appliedvar{x}{\emptyseq}$, which are included in values.

The \emph{bindings} in the language are the following. Variable $x$ is bound in $M$ in
terms $\abstraction{x}{}{M}$ and $\letbox xNM$; variables $x$ and $y$ are bound
in $M$ in term $\pairDestructor xyNM$; the variables in $\overline x$ are bound
in $M$ in contextual value $\ctxval{\overline x}M$.
The set of bound and free variables in terms ($\fv M$) are defined accordingly.
We follow the variable convention whereby all bound variables are chosen to be
different from the free variables,
in all contexts~\cite{DBLP:books/daglib/0067558,DBLP:books/daglib/0005958}.

\emph{Substitution} is defined accordingly, from the bindings in the language,
while taking advantage of the variable convention.
Substitution is rather conventional except for the fact that the term language
includes applied modal variables $\appliedvar x{\overline \sigma}$ rather than
ordinary variables $x$.
We denote by $\substitution{\sigma}{x}M$ the term obtained by replacing the free
occurrences of variable $x$ by the contextual value $\sigma$ in term $M$, and
similarly for $\substitution{\sigma}{x}\rho$.
We detail the novelties.
%
\begin{alignat*}{2}
&\substitution{\ctxval{\overline z}M}{x} (\appliedvar x{\overline\rho}) = \substitution{\substitution{\ctxval{\overline z}M}{x}\overline\rho}{\overline z} M
\qquad
&\substitution \sigma x (\appliedvar y{\overline\rho}) = \appliedvar y {\substitution{\sigma}{x}{\overline\rho}} \quad\text{if } x \neq y
  %
  \\[1ex]
&\substitution \sigma x {(\ctxval{\overline y}M)} = \ctxval{\overline y}{\substitution \sigma x M}
\end{alignat*}
Multiple substitution $\substitution{\overline\rho}{\overline z}M$ is defined only when
$\overline \rho$ and $\overline z$ are sequences of the same length. The variable
convention ensures that the variables in $\overline z$ are pairwise distinct and
not free outside $M$, hence the substitution of the various variables in
$\overline z$ can be performed in sequence.
Substitution on applied variables,
$\substitution{\ctxval{\overline z}{M}}{x}(\appliedvar{x}{\overline\rho})$,
triggers further substitution for $\overline z$ in $M$.
For example, $\substitution{\ctxval{z_1,z_2}{\application{\application{\send}{z_1}}{z_2}}} x ({\appliedvar x{y,42}})
= \substitution{y}{z_1} \substitution{42}{z_2} (\application{\application{\send}{z_1}}{z_2})
= \application{\application{\send}{y}}{42}$.
Substitution on the remaining term constructors is a homomorphism; for example
$\substitution\sigma x (\cbox \rho) = \cbox{(\substitution \sigma x \rho)}$
and
$\substitution \sigma x (\letbox yMN) = \letbox y{\substitution \sigma xM}{\substitution \sigma xN}$.
Since substitution for an applied variable triggers another, well-definedness is
not trivial. We show that substitution is well defined for typable terms
(provided that a few additional side conditions are met) in \cref{sec:proofs}.

If terms provide for the sequential part of the language, \emph{processes} deal
with concurrency and are used as a runtime only. Programmers write terms $M$
that are run on an initial thread $\thread M$ that may eventually fork new
threads and create new communication channels. Process $P\PAR Q$ denotes the
parallel composition of two processes. Process $\channelBinding xyP$ introduces
in process $P$ a communication channel described by its two endpoints $x$ and
$y$.
Our development is quite standard in the
literature~\cite{DBLP:journals/jfp/GayV10}.


\begin{figure*}[t!]
\begin{mathpar}
\inferrule* [lab=\ruletrbeta]
{}
{\reduction{\application{(\abstraction{x}{M})}{v}}{\substitution{\ctxval{\emptyseq}v}{x}M}}

\inferrule* [lab=\ruletrsplit]
{}
{\reduction{\pairDestructor{x}{y}{\pair{u}{v}}{M}}{
    \substitution{\ctxval{\emptyseq}{u}}{x} \substitution{\ctxval{\emptyseq}{v}}{y} M}}

\inferrule* [lab=\ruletrletbox]
{}
{\reduction{\letbox x {\cbox\sigma} M}{\substitution{\sigma}{x}M}}

\inferrule* [lab=\ruletrfix]
{}
{\reduction{\application \fix
    v}{\application{v}{(\abstraction{x}{\application{(\application \fix v)}{x}}})}}

%
\inferrule* [lab=\ruletrctx]
{\reduction{M}{N}}
{\reduction{\evaluationContextApply{\evaluationContextVar}{M}}{\evaluationContextApply{\evaluationContextVar}{N}}}
\end{mathpar}
\caption{Term evaluation \fbox{$\reduction{M}{M}$}}
\label{figure:term-evaluation}
\end{figure*}%


%
\emph{Evaluation} on terms is given by the relation $\reduction{M}{N}$, defined
by the rules in \cref{figure:term-evaluation}. It includes function application
(\ruletrbeta), the elimination of a pair of values $\pair uv$ and their binding
to variables $x$ and $y$ in $M$ (\ruletrsplit).  Notice that substitution is
defined for contextual values only, hence we lift value $u$ to $\ctxval
\emptyseq u$, and similarly for $v$.
Rule \ruletrletbox unboxes a boxed code fragment $\sigma$, binds it to $x$ to be
used in $M$.
Rule \ruletrfix unfolds the call by value fixed point construtor \fix~\cite{DBLP:books/daglib/0085577}.
%
Finally, rule \ruletrctx evaluates terms under evaluation contexts, the syntax
of which is in \cref{figure:expressions-processes}. These are the
standard call-by-value evaluation contexts. Notice we do not allow
reduction inside boxes just as we do not allow reduction under $\lambda$.




\begin{figure*}[t!]
\begin{mathpar}
\inferrule* []
{}
{\processCong
  {P}
  {\parallelComposition{\thread{\unit}}{P}}}

\inferrule* []
{}
{\processCong
  {\parallelComposition{P}{Q}}
  {\parallelComposition{Q}{P}}}

\inferrule* []
{}
{\processCong
  {\parallelComposition{(\parallelComposition{P_1}{P_2})}{P_3}}
  {\parallelComposition{P_1}{(\parallelComposition{P_2}{P_3})}}}

\inferrule* []
{}
{\processCong
  {\parallelComposition{\channelBinding{x}{y}{P}}{Q}}
  {\channelBinding{x}{y}{(\parallelComposition{P}{Q})}}}


\inferrule* []
{}
{\processCong
  {\channelBinding{x}{y}{P}}
  {\channelBinding{y}{x}{P}}}

\inferrule* []
{}
{\processCong
  {\channelBinding{x}{y}{\channelBinding{z}{w}{P}}}
  {\channelBinding{z}{w}{\channelBinding{x}{y}{P}}}}
\end{mathpar}
\caption{Structural congruence \fbox{$\processCong{P}{Q}$}}
\label{figure:structural-congruence}
\end{figure*}


\emph{Reduction} on processes is given by the the relation $\reduction PQ$,
defined by the rules in \cref{figure:process-reduction}. As customary in the
$\pi$-calculus, reduction builds on a further relation---structural congruence
$P \equiv Q$---that provides for the syntactic rearrangement of processes, while
preparing these for reduction~\cite{DBLP:journals/mscs/Milner92}.

\emph{Structural congruence} is the least congruence relation generated by the
axioms in \cref{figure:structural-congruence}. The first axiom posits thread
$\thread\unit$ as the neutral element of parallel composition. The next two
axioms tell that parallel composition is commutative and associative. Law
$\channelBinding xy P \PAR Q \equiv \channelBinding x y {(P \PAR Q)}$ is called
scope extrusion and allows the scope of a $\nu$-binder to expand to a new
process $Q$ or to retract from it, as needed.
Further axioms allow exchanging the order of the two endpoints of a channel, and
exchanging the order of $\nu$-binders.
In scope extrusion, because of the variable convention, the condition `$x,y$ not
free in $Q$' is redundant: $x$ and $y$ occur bound in $\channelBinding xyP$ and
therefore cannot occur free in~$Q$.

\begin{figure*}[t!]
\begin{mathpar}
\inferrule* [lab=\ruleprexp]
{\reduction{M}{N}}
{\processReduction{\thread{M}}{\thread{N}}}

\inferrule* [lab=\ruleprfork]
{}
{\processReduction{\thread{\evaluationContextApply{E}{\application{\fork}{v}}}}{\parallelComposition{\thread{\evaluationContextApply{E}{\unit}}}{\thread{\application{v}{\unit}}}}}

\inferrule* [lab=\ruleprnew]
{}
{\processReduction{\thread{\evaluationContextApply{E}{\new}}}{\channelBinding{x}{y} {\thread{\evaluationContextApply{E}{\pair{x}{y}}}}}}

\inferrule* [lab=\ruleprclose]
{}
{\processReduction
  {\channelBinding{x}{y}
    {(\parallelComposition
      {\thread{\evaluationContextApply{E}{\application{\close}{x}}}}
      {\thread{\evaluationContextApply{F}{\application{\wait}{y}}}})}}
    {\parallelComposition
      {\thread{\evaluationContextApply{E}{\unit}}}
      {\thread{\evaluationContextApply{F}{\unit}}}}}

\inferrule* [lab=\ruleprcom]
{}
{\processReduction
  {\channelBinding{x}{y}
    {(\parallelComposition
      {\thread{\evaluationContextApply{E}{\application{\application{\send}{v}}{x}}}}
      {\thread{\evaluationContextApply{F}{\application{\receive}{y}}}})}}
  {\channelBinding{x}{y}
    {\parallelComposition
      {\thread{\evaluationContextApply{E}{x}}}
      {\thread{\evaluationContextApply{F}{\pair{v}{y}}}}}}}

\inferrule* [lab=\ruleprchoice]
{}
{\processReduction
  {\channelBinding{x}{y}
    {(\parallelComposition
      {\thread{\evaluationContextApply{E}{\application{\select{l'}}{x}}}}
      {\thread{\evaluationContextApply{F}{\match{y}{l \rightarrow M_l} {l \in L}}}})}}
  {\channelBinding{x}{y}
    {(\parallelComposition
      {\thread{\evaluationContextApply{E}{x}}}
      {\thread{\evaluationContextApply{F}{\application{M_{l'}}{y}}}})}}}

\inferrule* [lab=\ruleprpar]
{\processReduction{P}{P'}}
{\processReduction{\parallelComposition{P}{Q}}{\parallelComposition{P'}{Q}}}

\inferrule* [lab=\ruleprbind]
{\processReduction{P}{Q}}
{\processReduction{\channelBinding{x}{y}{P}}{\channelBinding{x}{y}{Q}}}

\inferrule* [lab=\ruleprcong]
{\processCong{P}{P'} \\ \processReduction{P'}{Q'} \\ \processCong{Q'}{Q}}
{\processReduction{P}{Q}}
\end{mathpar}
\caption{Process reduction \fbox{$\reduction{P}{P}$}}
\label{figure:process-reduction}
\end{figure*}


We now describe the reduction rules. Rule \ruleprexp lifts term evaluation to
process reduction. Rule \ruleprfork creates a new thread. Value $v$ is
supposed to be a suspended computation (a thunk), so that $\application v \unit$
runs the computation. The hole in the context in the original thread is filled
with \unit: all communication between the two threads must be accomplished via
message passing on channels free in $v$. Rule \ruleprnew creates a new channel
denoted by its two endpoints $x$ and $y$. The hole in the thread is filled with
a pair $\pair xy$ so that the thread may manipulate the channel.

The next three rules manipulate channels via their endpoints $x,y$. For ease of
reading we abbreviate contextual term value $\appliedvar x \emptyseq$ to $x$,
and similarly for $y$. Rule \ruleprclose closes a channel: one thread must be
closing one endpoint, the other waiting for the channel to be closed. The
$\nu$-binder is eliminated since the channel cannot be further used. Rule
\ruleprcom exchanges a value between a \send and a \receive thread. The
result of sending is the endpoint itself, $x$; the result of receiving is a pair
composed of the value exchanged and the endpoint $y$. The two threads may then
continue exchanging values on these endpoints. Rule \ruleprchoice exercises the
choice of a branch in a \matchk process. The label selected on a thread ($l'$)
chooses branch $M_{l'}$ on the other thread. The result of selection is the
endpoint itself, as in \send; the result of match is $M_{l'}$, which is supposed to
be a function, applied to endpoint $y$. Similarly to \ruleprcom, the two threads
may continue interacting on channel $xy$.
Finally, the last three rules allow reduction under parallel composition and
$\nu$-binders, and incorporate structural congruence in reduction.



\section{Types and typing assignment}
\label{section:types}

This section introduces the notion of type and a type assignment system to
terms and to processes.


We build the language of types on the syntactic category of \emph{type
  references}, $a, b$. We further use letters $k,n$ to denote natural numbers,
and may call them \emph{levels}. \begin{figure*}[t!]
  \begin{alignat*}{4}
    &\text{Multiplicity} &&\multOp  && \grammarEq  &&
    \multlin \grammarOr \multun
    \\
    &\text{Type}  &&T, U && \grammarEq  &&
    \Unit \grammarOr
    \functionType{T}{\multOp}{T} \grammarOr
    \productType{T}{T} \grammarOr
    \contextualModalType{\tau^{n+1} } \grammarOr
    S
    \\
    &\text{Session type}  &&R, S  && \grammarEq  &&
    \Close \grammarOr \Wait \grammarOr
    \modeType{\sendOp}{T}{S} \grammarOr
    \modeType{\receiveOp}{T}{S} \grammarOr
    \choiceType{\internalChoiceOp}{l\colon S_l}{l \in L} \grammarOr
    \choiceType{\externalChoiceOp}{l\colon S_l}{l \in L} \grammarOr
    a \grammarOr \recursiveType{a}{S}
    \\
    &\text{Contextual type } &&\tau^n && \grammarEq  &&
    \contextualType{\overline \tau}{n}{T}
    \\
    &\text{Typing context}  &&\Gamma, \Delta  && \grammarEq  &&
    \emptyseq \grammarOr
    \Gamma, \binding{x}{\tau}
  \end{alignat*}
  \caption{Types}
  \label{figure:types}
\end{figure*}
%
%
The syntax of \emph{types} is in \cref{figure:types}. \emph{Multiplicities} $m$,
include $\multlin$ denoting a \emph{linear} resource---a resource that must be
used exactly once---and $\multun$ describing an \emph{unrestricted} resource---a
resource that may be used an unbounded number of times, including zero.

Type $\Unit$ (for constant $\unit$) can be thought as a placeholder for other
base types, such as those for integer or boolean values.
Further types include those for linear and unrestricted functions,
$\functionType{T}{\multlin}{U}$ and $\functionType{T}{\multun}{U}$, that for
linear pairs, $\productType TU$ (for simplicity we do not consider unrestricted
pairs), and for box types $\contextualModalType{\tau}$, that is,
code fragments characterised by contextual types $\tau$.
The grammar for contextual types is indexed by levels; when levels are not important
we omit them and abbreviate contextual types to $\tau$.

\emph{Session types} describe communication channel endpoints.
Types $\Close$ and $\Wait$ are for channels ready, or waiting, to be closed, respectively.
%
Type $\modeType{\sendOp}{T}{S}$ and $\modeType{\receiveOp}{T}{S}$ are for channels that send, or receive, values of type $T$ and continue as $S$, respectively.
Type $\choiceType{\internalChoiceOp}{l\colon S_l}{l \in L}$ is for channels that may select one branch $k\in L$ and proceed as $S_k$ (internal choice).
Its dual, $\choiceType{\externalChoiceOp}{l\colon S_l}{l \in L}$, is for channels that offer a menu of labelled choices (external choice).
Recursive session types are built from type references $a$ and recursion
$\recursiveType{a}{S}$. We take an \emph{equi-recursive} approach to types,
not distinguishing a recursive type from its unfolding~\cite{DBLP:conf/lics/AbadiF96,DBLP:books/daglib/0005958}.

\begin{figure*}[t!]
\begin{mathpar}
\inferrule* []
{}
{\dualRelation{\Close}{\Wait}}

\inferrule* []
{}
{\dualRelation{\Wait}{\Close}}

\inferrule* []
{\dualRelation{S}{R}}
{\dualRelation{\modeType{\sendOp}{T}{S}}{\modeType{\receiveOp}{T}{R}}}

\inferrule* []
{\dualRelation{S}{R}}
{\dualRelation{\modeType{\receiveOp}{T}{S}} {\modeType{\sendOp}{T}{R}}}

\inferrule* []
{\dualRelation{R}{S \substitution{\recursiveType{a}{S}}{a}}}
{\dualRelation{R}{\recursiveType{a}{S}}}

\inferrule* []
{\dualRelation{R \substitution{\recursiveType{a}{R}}{a}}{S}}
{\dualRelation{\recursiveType{a}{R}}{S}}

\inferrule* []
{\dualRelation{S_l}{R_l} \\ \forall l \in L}
{\dualRelation{\choiceType{\internalChoiceOp}{l : S_l}{l \in L}}{\choiceType{\externalChoiceOp}{l : R_l}{l \in L}}}

\inferrule* []
{\dualRelation{S_l}{R_l} \\ \forall l \in L}
{\dualRelation{\choiceType{\externalChoiceOp}{l : S_l}{l \in L}}{\choiceType{\internalChoiceOp}{l : R_l}{l \in L}}}
\end{mathpar}
\caption{Type duality (coinductive) \fbox{$\dualRelation SS$}}
\label{figure:type-duality}
\end{figure*}%


The notion of \emph{duality} in session types is captured by relation
$\dualRelation RS$, in \cref{figure:unrestricted-predicate}. Types
\Wait{} and \Close{}; input and output; and internal and external choice are dual to each other.
Recursive types are unfolded.
The definition is coinductive, so that we accept possible infinite derivations.
We refer to Gay \etal for details~\cite{DBLP:journals/corr/abs-2004-01322}.

\emph{Box types} are of the form
$\contextualType{\overline{\tau}}{n}{T}$, where $\overline\tau$ denotes a
possible empty sequence of contextual types.
Such a type represents a code fragment $M$ of type $T$, parameterised by a sequence
of variables $\overline x$ of contextual types $\overline\tau$, together called
contexts~\cite{DBLP:journals/pacmpl/JangGMP22,DBLP:conf/esop/MuraseNI23}.
Ordinary (modal) variables are typed at level 0.
%
The context of an ordinary variable is always empty, which justifies writing $T$
for the contextual type $\contextualType \emptyseq 0 T$.
Furthermore, in a type $\contextualType{\overline{\tau}}{n+1}{T}$, we require
all contextual types in $\overline\tau$ to be of levels smaller than or equal to $n$.
The types in $\overline{\tau}$ characterise the parameters $\overline x$ of the
contextual value $\ctxval{\overline x}M$ and thus must be typed at levels smaller
than that of the of value itself.
We capture these intuitions with a family of type formation predicates, $\ctype
T n$, defined by the rules below.
\begin{mathpar}
  \inferrule* []{}{
    \ctype{(\contextualType \emptyseq 0 T)}{0}
  }

  \inferrule* []{
    \ctype {\overline\tau} n
  }{
    \ctype{(\contextualType{\overline\tau}{n+1}{T})}{n+1}
  }

  \inferrule* []{
    \ctype \tau n
  }{
    \ctype{\tau}{n+1}
  }
\end{mathpar}
Henceforth we assume that all contextual types $\tau$ are well formed, that is,
that $\ctype \tau n$ holds for some $n$.
For example, object
$\contextualType{(\contextualType{\Wait}{1}{U})}{1}{T}$ cannot be deemed a
well formed type.
This formulation adheres to the development of M\oe bius and its kinding
system~\cite{DBLP:journals/pacmpl/JangGMP22}.




Finally, the \emph{box type} $\contextualModalType{\tau^{n+1}}$ represents a code
fragment of contextual type $\tau$. Boxed code fragments are typed at levels
starting at 1.

%
\begin{figure*}[t!]
\begin{mathpar}
\inferrule* []
{\un{T}}
{\un{(\contextualType{\emptyseq}{0}{T})}}

\inferrule*
{}
{\un{(\contextualType{\overline\tau}{n+1}{T})}}

\inferrule* []
{}
{\un{\Unit}}

\inferrule* []
{}
{\un{(\functionType{T}{\multun}{U})}}

\inferrule* []
{}
{\un{(\contextualModalType{\tau})}}
\end{mathpar}
\caption{Unrestricted types \fbox{$\un{T}$} \fbox{$\un{\tau}$}}
\label{figure:unrestricted-predicate}
\end{figure*}
%
%
%
Types denote resources that may be used exactly once or else an unbounded number
of types. The $\multun$ predicate is true of types and contextual types whose
values can be used an unbounded number of times. The definition is in
\cref{figure:unrestricted-predicate}.
Contextual types of level 1 or above are all unrestricted, whereas those of
level 0 are unrestricted only when the type $T$ of the code in unrestricted.
Level 0 represents ordinary code (of type $T$) and hence
$\contextualType{\emptyseq}{0}{T}$ is unrestricted if $T$ is. On the other hand,
levels higher than 0 represent boxed code fragments that can be duplicated or
discarded at will, hence they are all of unrestricted nature.
For types, \Unit, unrestricted arrows $\functionType{T}{\multun}{U}$ and
box types $\contextualModalType{\tau}$ are the only unrestricted
types.
The rules in \cref{figure:unrestricted-predicate} are algorithmic. The $\multun$
predicate is thus decidable, which allows us to talk of its negation. A
contextual type $\tau$ that is not unrestricted is called linear. The
corresponding predicate is denoted $\lin\tau$.
Two further predicates describe upper and lower bounds to the levels of
contextual types. The predicate $<k$ is true of contextual types
$\contextualType{\overline\tau}{n}{T}$ such that $n<k$. The predicate $\ge k$ is
true of contextual types $\contextualType{\overline\tau}{n}{T}$ such that
$n \ge k$.

\emph{Typing contexts} $\Gamma,\Delta$ bind variables $x$ to contextual types
$\tau$. We assume no variable is bound twice in the
same context and take typing contexts up to the \emph{exchange} of individual
entries. The remaining two substructural rules---contraction and weakening---are
handled by context split (described below) and by allowing contexts with
unrestricted types in axioms.
%
The four predicates are then lifted to typing contexts in the standard way. For
example $\un\Gamma$ is true of contexts
$\binding{x_1}{\tau_1}, \dots, \binding{x_k}{\tau_k}$ such that $\un{\tau_1}$,
\dots, $\un{\tau_k}$.
Intuitively, $\Gamma < n$ is true of a context $\Gamma$ suitable for the
\emph{local variables} in a code fragment of level $n$, whereas $\Gamma \ge n$
is true of a context suitable for the \emph{outer variables}. Notice that there
is no context $\Gamma$ for which $\Gamma < 0$ holds, with the exception of the
empty context $\emptyseq$. On the other hand, $\ctxouter\Gamma 0$ is a tautology.

We use notation $\ctxun\tau$ to denote a contextual type $\tau$ for which $\un\tau$ holds, and similarly for $\ctxlin\tau$.
For contexts, we use notation $\ctxun\Gamma$ to denote a context $\Gamma$ for which $\un\Gamma$ holds, and similarly for $\ctxlocal\Gamma n$ and $\ctxouter\Gamma n$.

\begin{figure*}[t!]
\begin{align*}
\typeof{\close} &= \functionType{\Close}{\multun}{\Unit}\\
\typeof{\wait} &= \functionType{\Wait}{\multun}{\Unit}\\
\typeof{\send} &= \functionType{T}{\multun}{\functionType{\modeType{\sendOp}{T}{S}}{\multlin}{S}}\\
\typeof{\receive} &= \functionType{\modeType{\receiveOp}{T}{S}}{\multun}{\productType{T}{S}}\\
\typeof{\select{k}} &= \functionType{\choiceType{\internalChoiceOp}{l : S_l}{l \in L}}{\multun}{S_k}, \text{with } k \in L\\
\typeof{\new} &= \functionType{\Unit}{\multun}{\productType{R}{S}}, \text{with } \dualRelation{R}{S}\\
\typeof{\fork} &= \functionType{(\functionType{\Unit}{\multlin}{\Unit})}{\multun}{\Unit}\\
\typeof{\fix} &= \functionType{(\functionType{(\functionType{T}{\multun}{T})}{\multun}{(\functionType{T}{\multun}{T})})}{\multun}{(\functionType{T}{\multun}{T})}\\
\typeof{\unit} &= \Unit
\end{align*}
\caption{Type schemes for constants \fbox{$\typeof{c} = T$}}
\label{figure:types-schemes-constants}
\end{figure*}

%
%
The \emph{types for constants} are described by the type schemes in
\cref{figure:types-schemes-constants}. Those for session types are taken mostly
from Gay and Vasconcelos~\cite{DBLP:journals/jfp/GayV10}. Constants \close{} and
\wait{} close different ends of a channel and return \Unit. Constant \send
accepts a value and a channel on which to send the value and returns the
continuation channel; the second arrow is linear for $T$ may denote a linear
value. Constant \receive{} accepts a channel and returns a pair composed of the
value read and the continuation channel. Constant $\application \select k$
accepts an external choice type selects branch $k$, returning the appropriate
continuation. Constant \new{} denotes a suspended computation that, when invoked
(with \unit), returns a pair of channel endpoints of dual types. Constant
\fork{} receives a suspended computation, spawns a new thread to run the
computation and returns \Unit. Finally the type of \fix is that of the
call-by-value lambda calculus fixed-point combinator; all arrows are
unrestricted for the argument function may be used an unbounded number of times
(in each recursive call), including zero (in the base case), and \fix itself can
be used an unbounded number of times.

There will be times when the type system must merge typing contexts coming from different subderivations into a single context.
\emph{Context split} (read bottom-up) does the job.
The rules are in \cref{figure:context-split} and taken verbatim from Walker~\cite{walker:substructural-type-systems}: linear types go left or right (but not in both directions); unrestricted types are copied into both contexts.

\begin{figure}[t!]
  \begin{mathpar}
    \inferrule*
    {}
    {\emptyseq = \ctxsplit \emptyseq \emptyseq}
    
    \inferrule*
    {\Gamma = \ctxsplit {\Gamma_1} {\Gamma_2}} 
    {\Gamma, \binding{x}{\ctxlin\tau} = \ctxsplit {(\Gamma_1, \binding{x}{\ctxlin\tau})} {\Gamma_2} }

    \inferrule*
    {\Gamma = \ctxsplit {\Gamma_1} {\Gamma_2}} 
    {\Gamma, \binding{x}{\ctxlin\tau} = \ctxsplit {\Gamma_1} {(\Gamma_2, \binding{x}{\ctxlin\tau})}}

    \inferrule*
    {\Gamma = \ctxsplit {\Gamma_1} \Gamma_2} 
    {\Gamma, \binding{x}{\ctxun\tau} = \ctxsplit {(\Gamma_1, \binding{x}{\ctxun\tau})} {(\Gamma_2, \binding{x}{\ctxun\tau})}}
  \end{mathpar}
  \caption{Context split \fbox{$\Gamma = \ctxsplit \Gamma \Gamma$}}
  \label{figure:context-split}
\end{figure}



Typing judgements are of the form $\isterm \Gamma M T$, stating that term $M$ has
type~$T$ under typing context $\Gamma$. The rules for the judgements, which we
describe below, are in \cref{figure:term-formation}.
\begin{figure*}[t!]
\begin{mathpar}
\inferrule* [lab=\ruletsconst]
{} 
{\isterm{\ctxun\Gamma}{\constant}{\typeof{\constant}}}

%

\inferrule* [lab=\ruletsvar] 
{\isterm{\overline\Gamma}{\overline\sigma}{\overline\tau}}
{\isterm{(\ctxun{\Delta}, \binding{x}{(\contextualType{\overline{\tau}}{n}{T})) \csplit \overline{\Gamma} 
  } }{\appliedvar{x}{\overline\sigma}}{T}}

%
\inferrule* [lab=\ruletsarrowlini]
{\isterm{\Gamma,\binding{x}{(\contextualType{\emptyseq}{0}{T})}}{M}{U}}
{\isterm{\Gamma}{\abstraction{x}{}{M}}{\functionType{T}{\multlin}{U}}}

\inferrule* [lab=\ruletsarrowuni]
{\isterm{\ctxun\Gamma,\binding{x}{(\contextualType{\emptyseq}{0}{T})}}{M}{U}} 
{\isterm{\ctxun\Gamma}{\abstraction{x}{}{M}}{\functionType{T}{\multun}{U}}}

\inferrule* [lab=\ruletsarrowe]
{\isterm{\Gamma}{M}{\functionType{T}{\multOp}{U}} \\\\ \isterm{\Delta}{N}{T}}
{\isterm{\Gamma \csplit \Delta}{\application MN}{U}}  

\inferrule* [lab=\ruletspairi]
{\isterm{\Gamma}{M}{T} \\ \isterm{\Delta}{N}{U}}
{\isterm{\Gamma \csplit \Delta}{\pair{M}{N}}{\productType{T}{U}}}

\inferrule* [lab=\ruletspaire]
{\isterm{\Gamma}{M}{\productType{T}{U}} \\\\ \isterm{\Delta,\binding{x}{(\contextualType \emptyseq 0 T)}, \binding{y}{(\contextualType\emptyseq 0 U)}}{N}{V}}
{\isterm{\Gamma \csplit \Delta}{\pairDestructor{x}{y}{M}{N}}{V}}

\inferrule* [lab=\ruletsctxt]
{\isterm{\ctxouter{\Gamma}{n},\ctxlocal{\overline{\binding{x}{\tau}}}{n}}{M}{T}}
{\isterm{\ctxouter{\Gamma}{n}}{\ctxval{\overline x}{M}}{(\contextualType{\overline \tau}{n}{T})}}

\inferrule* [lab=\ruletsboxi]
{\isterm{\ctxun\Gamma} \sigma \tau}
{\isterm{\ctxun\Gamma, \ctxun\Delta} {\cbox\sigma}{\contextualModalType{\tau}}}
%
%
%

\inferrule* [lab=\ruletsboxe]
{\isterm{\Gamma} {M}{\contextualModalType{\tau}} \\ \isterm{\Delta,\binding x {\tau}} {N}{T}}
{\isterm{{\Gamma} \csplit \Delta}{\letbox{x}{M}{N}}{T}}%

\inferrule* [lab=\ruletsmatch]
{\isterm {\Gamma} {M} {\choiceType {\externalChoiceOp} {l : S_l} {l \in L}} \\\\ \isterm {\Delta} {N_l} {\functionType{S_l}{\multlin}{T}} \\ (\forall l \in L)}
{\isterm {\Gamma \csplit \Delta} {\match {M} {l : N_l} {l \in L}} {T}}
\end{mathpar}
\caption{Term formation \fbox{$\isterm{\Gamma}{M}{T}$}}
\label{figure:term-formation}
\end{figure*}


\emph{Contextual terms} are closures of the form $\ctxval{\overline x}{M}$, closing
term $M$ over a sequence of variables $\overline x$. Such a contextual term is
given a contextual type $\contextualType{\overline \tau}{n}T$ when $T$ is of
type $M$ under the hypothesis that the variables in $\overline x$ are of the
types in $\overline \tau$ (rule \ruletsctxt).
%
%
Intuitively, the free variables of $M$ are split in two groups: local and outer.
The local variables, $\overline x$, are distinguished in the contextual term
$\ctxval{\overline x}M$ and typed under context
$\ctxlocal{\overline{\binding x \tau}} n$ where $n$ is an upper bound of the
levels $k_i$ assigned to each local variable $x_i$ in $\overline x$. The level
of each local variable is arbitrary, as long as it is smaller than $n$.
Outer variables are typed under context $\ctxouter \Gamma n$.
Natural number $n$ thus denotes the level at which the code fragment
$\ctxval{\overline x}M$ is typed. In general, a contextual term can be typed at
different levels. All resources (variables) in the context in the conclusion, as
well as resources $\overline x$ are consumed in the derivation of $M$.

\emph{Modal variables} are of the form $\appliedvar{x}{\sigma_1\cdots \sigma_k}$,
denoting a contextual term applied to contextual values
$\sigma_1\cdots \sigma_k$. To type each $\sigma_i$ we need a separate context $\Gamma_i$.
The type of $x$ must be a contextual type of the form $\contextualType{\tau_1\cdots \tau_k}{n}{T}$ and each contextual term $\sigma_i$ must be of type $\tau_i$.
Since $x$ may occur free in any of $\sigma_i$, we use context split to allow the
entry $\binding x {(\contextualType{\tau_1\cdots \tau_k}{n}{T})}$ to occur in
each $\Gamma_i$. In this case, the contextual type must be unrestricted.


In rule \ruletsvar, the lengths of all sequences---$\overline{\Gamma}$, $\overline{\sigma}$ and
$\overline{\tau}$---must coincide. When the sequences are empty, the rule becomes
the conventional axiom for variables. In fact rule \ruletsvar{} is the natural
generalization of the axiom in Walker's linear
$\lambda$-calculus~\cite{walker:substructural-type-systems}, obtained by writing
type $\contextualType{\emptyseq}{n}{T}$ in the context as $T$, and writing
modal variable $\appliedvar x \emptyseq$ as $x$, thus $\isterm{\ctxun\Delta,\binding{x}{(\contextualType \emptyseq n T)}}{\appliedvar{x}{\emptyseq}}{T}$ abbreviates to $\isterm{\ctxun\Delta,\binding{x}{T}}{x}{T}$.
%
%
%
%
The axiom justifies the presence of the unrestricted context $\Delta$ in the
conclusion of \ruletsvar.

We briefly pause the presentation of the type system to address substitution.
The \emph{substitution principle} in the presence of linear typing
is tricky because substitution can duplicate or discard terms~\cite{DBLP:journals/jfp/GayV10,walker:substructural-type-systems}.
In addition, we have to take levels of types into account.
The resulting substitution principle can be stated as follows.
\begin{mathpar}
  \inferrule*[lab=\rulesubs]
  {
    \isterm \Gamma \sigma \tau^n
    \\
    \isterm{\Delta,\binding x {\tau^n}}{M}{T}
    \\
    (\lin{\tau} \text{ or } \un{\Gamma})
    \\
    (\Gamma \geq n)
  }{
    \isterm{\Gamma \csplit \Delta}{\substitution\sigma x M}{T}
  }
\end{mathpar}

We write $\Gamma \csplit \Delta$ to denote
the context $\Theta$ such that $\Theta = \Gamma \csplit \Delta$. When we do so,
we assume that the context split operation is defined.
The first side condition, $\lin{\tau^n} \text{ or } \un{\Gamma}$, prevents
free variables of linear types in $\sigma$ from being duplicated or discarded
by substitution $\substitution \sigma x M$ in this way: if $\tau^n$ is linear, then
no duplication or discarding occurs and, if $\un{\Gamma}$, then
free variables are safe to be duplicated or discarded.  The second side condition,
$\Gamma \geq n$, means that the levels of free variables in $\sigma$
have to be equal to or greater than the level of $\sigma$.

If $n=0$, that is when $x$ is an ordinary variable, the condition
$\Gamma \geq n$ trivially holds. Although the second formula in the condition
does not hold in general, we can show that it holds for values (that is, if
$\sigma = \ctxval \emptyseq v$ for some $v$). We can thus obtain the conventional
(call-by-value) substitution principle:
\begin{mathpar}
\inferrule
{\isterm{\Gamma}{\ctxval\emptyseq v}{(\contextualType \emptyseq 0 U)} \\\\ \isterm{\Delta,\binding x {(\contextualType \emptyseq 0 U})}{M}{T}
}
{\isterm{\Gamma \csplit \Delta}{\substitution {\ctxval\emptyseq v}xM}{T}}    

\text{abbreviated to}
  
\inferrule
{\isterm{\Gamma}{v}{U} \\ \isterm{\Delta,\binding x{}{U}}{M}{T}
}
{\isterm{\Gamma \csplit \Delta}{\substitution vxM}{T}}    
\end{mathpar}

Returning to the type system, we address the conventional typing rules for \emph{$\lambda$-abstraction and application} in the linear lambda calculus. The
rules are those of Walker~\cite{walker:substructural-type-systems} with the
necessary adaptation of the introduction rules to account for contexts featuring
contextual types.
There are two introduction rules, one for the linear, the other for the
unrestricted arrow. In either case, the $\lambda$-bound variable $x$ is local to
$M$, hence of level $0$. We record this fact by assigning $x$ a type of the form
$\contextualType \emptyseq 0 T$. For the unrestricted arrow we require an
unrestricted context, given that the context contains entries for the free
variables in term $\abstraction xM$ and this, being unrestricted, may be
duplicated or discarded.

The \emph{box introduction and elimination} rules are as follows.
In rule \ruletsboxi, the type of the box is the type
$\contextualModalType\tau$ if the contextual term $\sigma$ has contextual type $\tau$.
The context for the box elimination rule, \ruletsboxe{}, is split into two: one
part ($\Gamma$) to type $M$, the other ($\Delta$) to type $N$. Term
$M$ must denote a boxed code fragment, hence the type of $M$ must be
$\contextualModalType\tau$. Term $N$ is typed under
context $\Delta$ extended with an entry for~$x$.

Linear type systems with a context split operator perform all the required
weakening at the leaves of derivations. Hence all axioms expect a $\ctxun\Gamma$
context to discard unrestricted variables. This is the case of rules
\ruletsconst and \ruletsvar (with $\overline\sigma$ empty), and is all we need
in most cases. But, when typing a box, we need weakening for a different reason:
to discard unrestricted resources of small level (of levels smaller than that
necessary to type the code fragment inside the box). Context $\ctxun\Delta$
plays this role; we expect this context to contain all unrestricted entries of
levels below~$n$. Linear resources cannot be discarded in any case.

Rule \ruletsboxi is perfectly aligned with preceding
work~\cite{DBLP:journals/jacm/DaviesP01,DBLP:journals/pacmpl/JangGMP22,DBLP:journals/jfp/NanevskiP05},
where level control is checked at box introduction. A derived rule, composed of
\ruletsctxt{} followed by \ruletsboxi, coincides with the box introduction rule
of M\oe bius~\cite{DBLP:journals/pacmpl/JangGMP22}:
\begin{mathpar}
  \inferrule{
    \isterm{\ctxun \Gamma,\ctxlocal{\overline{\binding{x}{\tau}}}{n}}{M}{T} \\ (\ctxun \Gamma \geq n)
  }{
    \isterm{\ctxun \Gamma, \ctxun\Delta}{\cbox{(\ctxval{\overline x}{M})}}{\contextualModalType{(\contextualType{\overline\tau}{n}{T})}}
  }
\end{mathpar}




\section{Main results}
\label{sec:type-safety}
This section discusses local soundness and completeness for the different type
constructors, as well as the type safety result
for the process language.

To ensure the introduction and elimination rules form a valid type constructor,
the local soundness and completeness conditions must be shown for that type
constructor \cite{DBLP:journals/jacm/DaviesP01,DBLP:journals/mscs/PfenningD01}.
Local soundness states that no extra information is gained by applying an
introduction rule directly followed by an elimination rule. Hence, this detour
should not be required to complete a derivation, since the same judgement can be
reached without it. Its dual, local completeness, states that we can recover all
the information present before applying an elimination rule, by simply applying
an introduction rule. Local soundness materialises the reduction rules, while
local completeness materialises the expansion rules, and the proof of local
soundness implies type preservation for the reduction rule.


We start by studying the rules for the introduction and elimination of the
arrow type.
For \emph{local soundness} we have:
%
\begin{mathpar}
  \inferrule*[right=\ruletsarrowe]{
    \inferrule*[right=\ruletsarrowlini]{
      \isterm{\Gamma,\binding x {(\contextualType{\emptyseq}{0}{T})}} M U
    }{
      \isterm{\Gamma}{\abstraction x{}M}{\functionType T \multun U}
    }
    \\
    \isterm \Delta v T
  }{
    \isterm{\Gamma \csplit \Delta}{\application{(\abstraction x{}M)}{v}}{U}
  }
  \quad\Rightarrow\quad
  \inferrule*[right=\rulesubs]{
    \inferrule*[right=\ruletsctxt]
    {\isterm{\Delta}{v}{T}}
    {\isterm{\Delta}{\ctxval \emptyseq v}{(\contextualType \emptyseq 0 T)}}
    \\
    \isterm{\Gamma,\binding x {(\contextualType{\emptyseq}{0}{T})}}{M}{U}
  }{
    \isterm{\Gamma \csplit \Delta}{\substitution {\ctxval{\emptyseq}{v}}xM}{U}
  }   
\end{mathpar}
The case for $\functionType{}{\multun}{}$, that is for rule \ruletsarrowuni against
\ruletsarrowe, is similar and can be obtained by adding the $\omega$ restriction
to context $\Gamma$.

For \emph{local completeness}, assume that $x$ is not free in $M$ and let $\tau$ be
the contextual type $\contextualType \emptyseq 0 T$. We must distinguish four cases,
according to when $\Gamma$ and/or $T$ are unrestricted or linear.
We start with the case when both $\Gamma$ and $T$ are linear. Then, $\tau$ is
linear and $\Gamma , \binding{x}{\tau} = \Gamma \csplit \binding{x}{\tau}$, which
justifies the instance of rule \ruletsarrowe below.
\begin{mathpar}
  \isterm{\Gamma} M {\functionType T \multlin U}
  \quad\Rightarrow\quad
  \inferrule*[right=\ruletsarrowlini]
  {
    \inferrule*[right=\ruletsarrowe]
    {
      \isterm{\Gamma} M {\functionType T \multlin U}
      \\
      \inferrule*[right=\ruletsvar]
      { }
      {\isterm{\binding{x}{\tau}} {\appliedvar x \emptyseq} T}
    }
    {
      \isterm{\Gamma,\binding{x}{\tau}} {\application M{\appliedvar x \emptyseq}} U
    }
  }
  {\isterm{\Gamma} {\abstraction x {} {(\application M{\appliedvar x \emptyseq})}} {\functionType T \multlin U}}
\end{mathpar}
When $T$ is unrestricted, then so is $\tau$ and we have
$\Gamma , \binding{x}{\tau} = (\Gamma , \binding{x}{\tau}) \csplit
\binding{x}{\tau}$. We use weakening (see appendix)
to add $\binding x\tau$ to the context of $M$. We show the case when $\Gamma$ is
also unrestricted; the two remaining cases are similar.
\begin{mathpar}
  \isterm{\ctxun{\Gamma}} M {\functionType T \multun U}
  \quad\Rightarrow\quad
  \inferrule*[right=\ruletsarrowuni]
  {
    \inferrule*[right=\ruletsarrowe]
    {
      \inferrule*[right=\ruleweak]
      {
        \isterm{\ctxun{\Gamma}} M {\functionType T \multun U}
      }{
        \isterm{\ctxun{\Gamma}, \binding x\tau} M {\functionType T \multun U}
      }
      \and
      \inferrule*[right=\ruletsvar]
      { }
      {\isterm{\binding{x}{\tau}} {\appliedvar x \emptyseq} T}
    }
    {
      \isterm{\ctxun{\Gamma},\binding{x}{\tau}} {\application M{\appliedvar x \emptyseq}} U
    }
  }
  {\isterm{\ctxun{\Gamma}} {\abstraction x {} {(\application M{\appliedvar x \emptyseq})}} {\functionType T \multun U}}
\end{mathpar}


We now address the rules for box introduction and elimination. For \emph{local
  soundness} we have the following derivations, where rule \ruleweak introduces
unrestricted entries in the context.
\begin{mathpar}
\inferrule*[right=\ruletsboxe]
{
  \inferrule*[right=\ruletsboxi]
  {\inferrule*[right=\ruletsctxt]
    {\cdots\\\ctxun\Gamma \geq n}{\isterm{\ctxun {\Gamma}} \sigma \tau^n}}
    {\isterm{\ctxun\Gamma,\ctxun\Delta} {\cbox\sigma}{\contextualModalType{\tau^n}}}
    \\
    \isterm{\Theta,\binding x {\tau^n}} {N}{T}
}
{\isterm{(\ctxun\Gamma,\ctxun\Delta) \csplit \Theta}{\letbox{x}{\cbox\sigma}{N}}{T}}
\quad\Rightarrow\quad
\inferrule*[right=\rulesubs]{
  \inferrule*[right=\ruleweak]{
    \isterm{\ctxun\Gamma}{\sigma}{\tau^n}    
  }{
    \isterm{\ctxun\Gamma,\ctxun\Delta}{\sigma}{\tau^n}
  }    
  \\
  \isterm{\Theta,\binding x{\tau^n}}{M}{T}
  \\
  (\ctxun\Gamma \geq n)
}{
  \isterm{(\ctxun\Gamma , \ctxun\Delta) \csplit \Theta}{\substitution \sigma xN}{T}
}
\end{mathpar}

For \emph{local completeness} we first define eta expansion for contextual
types, written as $\etactx{x}{\tau}$ by the following rules, where variables in
$\overline y$ are taken freshly.
\begin{mathpar}
  \etactx{x}{\contextualType{\emptyseq}{0}{T}} = \ctxval{\emptyseq}{\appliedvar{x}{\epsilon}}

  \etactx{x}{\contextualType{\overline{\tau}}{n+1}{T}} = \ctxval{\overline{y}}{\appliedvar{x}{\overline{\etactx{y}{\tau}}}}
\end{mathpar}
Eta expansion $\etactx{x}{\tau}$ is defined at any contextual type $\tau$
because the level of $\tau$ strictly decreases at each step. We can confirm that
the following lemma holds; the proof is by induction on the level of $\tau$ and
can be found in the appendix.
\begin{lemma}
  \label{lem:etactx}
  $\isterm{x:\tau}{\etactx{x}{\tau}}{\tau}$.
\end{lemma}
%

We now witness local completeness.
Let $\Delta$ be the unrestricted part of $\Gamma$.
Then $\Gamma = \Gamma \csplit \Delta$. Let $\overline\rho$ be the sequence of contextual types, whose level is lower than $n$, and $\tau$ be the contextual type $\contextualType{\overline{\rho}} {n} T$.
We have $\ctxlocal{(\overline{\binding x \rho})}{n}$ and $\ctxouter{(u:\tau)}{n}$.
Furthermore, we can see the context $\ctxouter{(\binding u\tau)}{n}$ is
unrestricted because $n > 0$ is required by the well-formedness condition of
$\contextualModalType{\tau}$. We then have:
\begin{mathpar}
\isterm \Gamma M {\contextualModalType\tau}
\quad\Rightarrow\quad
\begin{prooftree}
  \isterm {\Gamma} M {\contextualModalType{\tau}} \quad
  \[
    \[
      \[
        \[
          \justifies
          \isterm{\overline{\binding{x}{\rho}}}{\overline{\etactx{x}{\rho}}}{\overline{\rho}}
          \using{\text{\cref{lem:etactx}}}
        \]
        \justifies
        \isterm {\overline{\binding{x}{\rho}} \csplit \binding u \tau
        }
        {\appliedvar{u}{\overline{\etactx{x}{\rho}}}}
        {T}
        \using \ruletsvar
      \]
      \justifies
      \isterm{\binding u \tau} {\ctxval{\overline
          x}{\appliedvar{u}{\overline{\etactx{x}{\rho}}}}} {\tau}
      \using \ruletsctxt
    \]
    \justifies
    \isterm{\Delta, \binding u \tau}{\cbox{(\ctxval{\overline
          x}{\appliedvar{u}{\overline{\etactx{x}{\rho}}}})}}{\contextualModalType\tau} \using \ruletsboxi
  \]
  \justifies
  \isterm{\Gamma \csplit
    \Delta}{\letbox{u}{M}{\cbox{(\ctxval{\overline
          x}{\appliedvar{u}{\overline{\etactx{x}{\rho}}}}})}}{\contextualModalType\tau}
  \using \ruletsboxe
\end{prooftree}
\end{mathpar}




\begin{figure*}[t!]
\begin{mathpar}
\inferrule* [lab=\rulepfexp]
{\isterm{\Gamma}{M}{\Unit}}
{\isprocess{\Gamma}{\thread{M}}}

\inferrule* [lab=\rulepfpar]
{\isprocess{\Gamma}{P} \\ \isprocess{\Delta}{Q}}
{\isprocess{\Gamma \csplit \Delta}{\parallelComposition{P}{Q}}}

\inferrule* [lab=\rulepfbind]
{\isprocess{\Gamma, \binding{x}{(\contextualType \emptyseq 0 R)}, \binding{y}{(\contextualType \emptyseq 0 S)}}{P} \\ \dualRelation{R}{S}}
{\isprocess{\Gamma}{\channelBinding{x}{y}{P}}}
\end{mathpar}
\caption{Process formation \fbox{$\isprocess{\Gamma}{P}$}}
\label{figure:process-formation}
\end{figure*}


The second part of this section addresses type safety.
Even if programmers are not supposed to write processes directly, we still need
to type them in order to state and prove the type safety result for the process
language. The rules, in \cref{figure:process-formation}, are straightforward and
taken from Vasconcelos~\cite{DBLP:journals/iandc/Vasconcelos12} and Thiemann
\etal~\cite{DBLP:journals/iandc/AlmeidaMTV22,DBLP:conf/icfp/ThiemannV16}.
In a thread $\thread M$, term $M$ must be of \Unit{} type. Given that the result
of the evaluation is discarded, any unrestricted type would do. The parallel
composition splits the context in the conclusion and uses one part for each
process. Scope restriction introduces entries for each of the two channel ends.
These must be of dual types.


We show a conventional safety property for the term language based on
\emph{preservation} and \emph{progress}---\cref{thm:preservation-terms,thm:progress}---following Pierce~\cite{DBLP:books/daglib/0005958}.
In the case of progress we take into consideration terms that may be
(temporarily) stuck waiting for a communication on a channel.


\begin{theorem}[Preservation]
\label{thm:preservation-terms}
\label{thm:preservation-processes}
\begin{enumerate}
\item If $\reduction MN$ and $\isterm{\Gamma}{M}{T}$, then
  $\isterm{\Gamma}{N}{T}$.
\item If $\isprocess{\Gamma}{P}$ and $\reduction PQ$, then $\isprocess{\Gamma}{Q}$.
\end{enumerate}
\end{theorem}


Progress is usually stated for closed terms. Because channel endpoints are also
variables, we must take into account terms with free endpoints, that is
variables of a session type. We say that a context $\Gamma$ \emph{contains
  session types only}, and write $\Gamma^S$ when $\Gamma$ is of the form
$\binding{x_1}{\contextualType \emptyseq
  {0}{S_1}},\dots,\binding{x_n}{\contextualType \emptyseq {0} {S_n}}$.
%

\begin{theorem}[Progress for evaluation]
  \label{thm:progress}
  Let $\isterm{\Gamma^S}{M}{T}$. Then,
  \begin{enumerate}
  \item\label{it:value} $M$ is a value, or
  \item $\reduction MN$, for some $N$, or
  \item $M$ is of the form $\evaluationContextApply EN$, for some $E$, with $N$
    of one of the following forms: $\close\ x$, $\wait\ x$, $\send\ v\ x$,
    $\receive\ x$, $\select\ l\ x$, $\match{x}{l \rightarrow N_l}{l \in L}$,
    $\application\new v$ or $\application \fork v$. 
  \end{enumerate}
\end{theorem}
%


%
The process language does not enjoy progress due to possible occurrences of
deadlocks. Safety for the process language is based on \emph{preservation} and
\emph{absence of runtime
  errors}---\cref{thm:preservation-processes,thm:absence-runtime-errors}---following
Honda \etal~\cite{DBLP:conf/esop/HondaVK98}.
In order to address the absence of runtime errors for the process language, we
start with a few definitions. The \emph{subject} of a term
$M$ 
is variable $x$ (denoting a channel endpoint) in the following cases and
undefined in all other cases.
\begin{mathpar}
  \application \close x

  \application \wait x

  \application{\send}{\application{v}{x}}

  \application \receive x

  \application {\select{k}} x

  \match{x}{l \rightarrow N_l}{l \in L}
\end{mathpar}

A process $\channelBinding{x}{y}{(\thread{\evaluationContextApply EM} \PAR \thread{\evaluationContextApply FN})}$ is a \emph{redex} if
\begin{itemize}
\item $M$ is $\application{\close}{x}$ and $N$ is $\application{\wait}{y}$,
\item $M$ is $\application{\send}{\application{v}{x}}$ and $N$ is $\application{\receive}{y}$,
\item $M$ is $\application{\select{l'}}{x}$ and $N$ is $\match{y}{l \rightarrow
    N_l}{l \in L}$ and $l'\in L$.
\end{itemize}
or conversely for processes $\channelBinding{x}{y}{(\thread{\evaluationContextApply FN} \PAR  \thread{\evaluationContextApply EM})}$.
This definition corresponds to the axioms \ruleprclose, \ruleprcom and \ruleprchoice in the operational semantics (\cref{figure:process-reduction}).
Saying that $P$ is a redex is equivalent to say that $P$ reduces by one of these rules.

Finally a process $P$ is a \emph{runtime error} if it is structural congruent to a process of the form $\channelBinding{x_1}{y_1}{\cdots\channelBinding{x_n}{y_n}{(P_1 \PAR \cdots \PAR P_m)}}$ and there are $i,j,k$ such that the subject of $P_i$ is $x_k$, the subject of $P_j$ is $y_k$ but $\channelBinding{x_k}{y_k}{(P_i \PAR P_j)}$ is not a redex.

\begin{theorem}[Absence of immediate errors]
\label{thm:absence-runtime-errors}
If $\isprocess{}{P}$ then $P$ is not a runtime error.
\end{theorem}


\section{Algorithmic type checking}
\label{section:algorithmic-type-checking}

The typing rules of \cref{figure:term-formation} are not
algorithmic. 
Two problems arise: guessing the right context split (a non-determistic procedure) and guessing the
types (and levels) of $\lambda$-abstractions and contextual values' bound variables.
For the former, we refactor the typing rules so they return the
unused part of the incoming context~\cite{walker:substructural-type-systems}.
For the latter, we work with an explicitly typed term language: we now write
$\constant^\tau$ to resolve the polymorphism of constants,
$\absT{\multOp}{x}{T}{M}$ to introduce the type of bound variable $x$, and
$\ctxval{(\overline{\binding{x}{\tau}})^n}{M}$ to introduce the contextual types
and the level of the contextual value.


\begin{figure*}[t!]
\begin{mathpar}
\inferrule* [lab=\ruleatconst]
{}
{\typesynth{\Gamma}{\constant^\tau}{\tau}{\Gamma}}

\inferrule* [lab=\ruleatvarun]
{\typesynth{\Gamma_1, \binding{x}{(\contextualType{\overline{\tau}}{n}{T})}}{\sigma_1}{\tau_1}{\Gamma_2} \\\\ \typesynth{\Gamma_2}{\sigma_2}{\tau_2}{\Gamma_3}\ \ldots\ \typesynth{\Gamma_k}{\sigma_n}{\tau_n}{\Gamma_{k+1}}}
{\typesynth{\Gamma_1, \ctxun{\binding{x}{(\contextualType{\overline{\tau}}{n}{T})}}}{\appliedvar{x}{\sigma_1, \ldots, \sigma_k}}{T}{\Gamma_{k+1}}}

\inferrule* [lab=\ruleatvarlin]
{\typesynth{\Gamma_1}{\sigma_1}{\tau_1}{\Gamma_2}\ \ldots\ \typesynth{\Gamma_k}{\sigma_k}{\tau_k}{\Gamma_{k+1}}}
{\typesynth{\Gamma_1, \binding{x}{(\contextualType{\overline{\tau}}{n}{T})}}{\appliedvar{x}{\sigma_1, \ldots, \sigma_k}}{T}{\Gamma_{k+1}}}

\inferrule* [lab=\ruleatarrowlini]
{\typesynth{\Gamma, \binding{x}{(\contextualType{\emptyseq}{0}{T})}}{M}{U}{\Delta}}
{\typesynth{\Gamma}{\absT{\multlin}{x}{T}{M}}{\functionType{T}{\multlin}{U}}{\ctxdiff{\Delta}{x}}}

\inferrule* [lab=\ruleatarrowuni]
{\typesynth{\Gamma,\binding{x}{(\contextualType{\emptyseq}{0}{T})}}{M}{U}{\Delta} \\\\ \ctxdiff{\Delta}{x} = \Gamma}
{\typesynth{\Gamma}{\absT{\multun}{x}{T}{M}}{\functionType{T}{\multun}{U}}{\Gamma}}

\inferrule* [lab=\ruleatarrowe]
{\typesynth{\Gamma_1}{M}{\functionType{T}{\multOp}{U}}{\Gamma_2} \\ \typesynth{\Gamma_2}{N}{T}{\Gamma_3}}
{\typesynth{\Gamma_1}{\application {M}N}{U}{\Gamma_3}}

\inferrule* [lab=\ruleatpairi]
{\typesynth{\Gamma_1}{M}{T}{\Gamma_2} \\ \typesynth{\Gamma_2}{N}{U}{\Gamma_3}}
{\typesynth{\Gamma_1}{\pair{M}{N}}{\productType{T}{U}}{\Gamma_3}}

\inferrule* [lab=\ruleatpaire]
{\typesynth{\Gamma_1}{M}{\productType{T}{U}}{\Gamma_2} \\\\ \typesynth{\Gamma_2, \binding{x}{(\contextualType \emptyseq 0 T)}, \binding{y}{(\contextualType{\emptyseq}{0}{U})}}{N}{V}{\Gamma_3}}
{\typesynth{\Gamma_1}{\pairDestructor{x}{y}{M}{N}}{V}{\ctxdiff{\ctxdiff{\Gamma_3}{x}}{y}}}

\inferrule* [lab=\ruleatboxi]
{\typesynth{\Gamma}{\sigma}{\tau}{\Gamma}}
{\typesynth{\Gamma}{\cbox\sigma}{\contextualModalType{\tau}}{\Gamma}}

\inferrule* [lab=\ruleatboxe]
{\typesynth{\Gamma_1}{M}{\contextualModalType{\tau}}{\Gamma_2} \\ \typesynth{\Gamma_2, \binding x {\tau}}{N}{T}{\Gamma_3}}
{\typesynth{\Gamma_1}{\letbox{x}{M}{N}}{T}{\ctxdiff{\Gamma_3}{x}}}

\inferrule* [lab=\ruleatmatch]
{\typesynth{\Gamma_1}{M}{\choiceType{\externalChoiceOp}{l : S_l}{l \in L}}{\Gamma_2} \\\\ \typesynth{\Gamma_2}{N_l}{\functionType{S_l}{\multlin}{T}}{\Gamma_3} \\ (\forall l \in L)}
{\typesynth{\Gamma_1}{\match{M}{l : N_l}{l \in L}}{T}{\Gamma_3}}

\inferrule* [lab=\ruleatctxt]
{\ctxdiv{\Gamma_1}{n}{\ctxouter{\Gamma_3}{n}, \ctxlocal{\Gamma_4}{n}} \\
  \Gamma_2 < n \\\\  \typesynth{\Gamma_2,\Gamma_3}{M}{T}{\Delta}}
{\typesynth{\Gamma_1}{(\ctxval{\Gamma_2^n}{M})}{(\contextualType{\overline \tau}{n}{T})}{\ctxdiff{\Delta}{\overline{x}}, \Gamma_4}}
\end{mathpar}
\caption{Algorithmic type checking \fbox{$\typesynth{\Gamma}{M}{T}{\Delta}$}}
\label{figure:algorithmic-type-checking}
\end{figure*}
%
%

We introduce a new type synthesis judgement
$\typesynth{\Gamma_{\text{in}}}{M_{\text{in}}}{T_{\text{ou}t}}{\Delta_{\text{out}}}$,
where context $\Gamma$ and term $M$ are considered inputs and type $T$ and
context $\Delta$ are considered outputs. The rules are in
\cref{figure:algorithmic-type-checking}. Some rules make use of \emph{context difference},
$\ctxdiff{\Gamma}{x} = \Delta$, an operation that ensures the binding for $x$ is
removed from $\Gamma$, in case the type bound is unrestricted, or it doesn't
even occur, in case the type is linear. We let operator $\ctxdiff{}{}$ be
left-associative.
\begin{mathpar}
\inferrule*
{\binding{x}{\tau} \not\in \Gamma}
{\ctxdiff{\Gamma}{x} = \Gamma}

\inferrule*
{ }
{\ctxdiff{(\Gamma, \binding{x}{\ctxun\tau}, \Delta)}{x} = \Gamma, \Delta}
\end{mathpar}

The typing rules are inspired by Almeida \etal~\cite{DBLP:journals/iandc/AlmeidaMTV22}.
In rules \ruleatconst, \ruleatarrowuni and \ruleatboxi, the input and output
contexts are the same, meaning no linear variables may be consumed.
Rules \ruleatarrowlini, \ruleatarrowuni, \ruleatpaire, \ruleatboxe and
\ruleatctxt ensure that bound variables do not escape their scope. In rule
\ruleatctxt, we must split the context ($\ctxdiv{\Gamma}{n}{\Delta}$) according
to the level $n$, using only the outer fragment to type the subexpression,
mimicking rule \ruletsctxt. The remaining rules are easy to understand.

If we denote by $M^\bullet$ an annotated term (as those in
\cref{figure:algorithmic-type-checking}), we can easily define an erasure
function, $\erase{M^\bullet}$ by induction on the structure of $M^\bullet$.
Then we have the following result.


\begin{theorem}[Algorithmic correctness]\
\label{theorem:algorithmic-soundness}
\label{theorem:algorithmic-completeness}
\begin{enumerate}
\item If $\typesynth{\Gamma}{M^\bullet}{T}{\Delta}$ and $\ctxun{\Delta}$, then $\isterm{\Gamma}{\erase{M^\bullet}}{T}$.
\item If $\isterm{\Gamma}{M}{T}$, then there is a $M^\bullet$ s.t.\ $M = \erase{M^\bullet}$, and $\typesynth{\Gamma}{M^\bullet}{T}{\Delta}$ and $\ctxun{\Delta}$.
\end{enumerate}
\end{theorem}



\section{Related work}
\label{sec:related}
Our approach to staged metaprogramming is based on contextual modal types.
The logical basis for this approach goes back to modal logic S4 \cite{prawitz1965natural,DBLP:journals/jacm/DaviesP01}, where a type $\Box T$ stands for a closed code fragment that evaluates to a value with type $T$.
Box types are then generalized by introducing contexts, such as $\Box (\overline{T_1} \vdash T_2)$, allowing free variables in a code fragment by listing their types in the box type.
These types are called contextual modal types~\cite{DBLP:journals/jfp/NanevskiP05,DBLP:journals/tocl/NanevskiPP08}.
To avoid confusion with contextual types, we call them box types in this paper.
M\oe bius~\cite{DBLP:journals/pacmpl/JangGMP22} further generalizes contextual modal types to hold contexts with different levels, allowing wider variety of metaprogramming.
There seems to be two styles when it comes to formalizing contextual modal types: one is a dual-context style~\cite{DBLP:journals/jacm/DaviesP01,DBLP:journals/jfp/NanevskiP05,DBLP:journals/tocl/NanevskiPP08} where box values are deconstructed with let-box syntax and meta-variables, and the other is Kripke-style or Fitch-style~\cite{DBLP:conf/fossacs/Clouston18,DBLP:journals/jacm/DaviesP01,DBLP:conf/esop/MuraseNI23,DBLP:journals/pacmpl/ValliappanRC22} where box values are deconstructed with an unquote term.
In this classification, M{\oe}bius \cite{DBLP:journals/pacmpl/JangGMP22} can be considered a calculus in dual-context style, but further generalized to allow multi-level contexts.
As discussed in the introduction, we chose M{\oe}bius as a basis 
because its syntax is more tractable than the Kripke/Fitch-style formulation.

Intensional analysis~\cite{DBLP:conf/esop/HuP24,DBLP:journals/pacmpl/ParreauxVSK18}, \ie the ability to perform pattern matching on code fragments, is a desirable feature with which to improve our work.
Despite the original formulation of M\oe bius also supporting intensional analysis, extending it to support linear types can be a non-trivial problem.
Supporting polymorphism is another potential extension.
M{\oe}bius supports generating code that depends on polymorphic types, which should be feasible in our type system.
Murase \etal proposed polymorphic contexts~\cite{DBLP:conf/esop/MuraseNI23} that can abstract several parts of a context by context variables. Such extension could also be useful in our type system, but it is not clear if we can extend our M{\oe}bius-based type system, given that polymorphic contexts as defined by Murase \etal~\cite{DBLP:conf/esop/MuraseNI23} are designed for a Fitch-style formulation.


Session types were introduced by Honda \etal~\cite{DBLP:conf/concur/Honda93,DBLP:conf/esop/HondaVK98,DBLP:conf/parle/TakeuchiHK94} in the decade of 1990.
The types we employ---input/output and internal/external choice---are from these early works.
The types for channels ready to be closed are from Caires \etal~\cite{DBLP:conf/concur/CairesP10,DBLP:journals/mscs/CairesPT16}.
The early works on session types were built on the $\pi$-calculus; session types for functional programming were later suggested by Gay and Vasconcelos~\cite{DBLP:journals/jfp/GayV10}.
The idea of using different identifiers to denote the two ends of a same communication channel is by Vasconcelos~\cite{DBLP:journals/iandc/Vasconcelos12}.

The only works integrating linear type systems and multi-stage programming that
we are aware of are those proposed by Georges
\etal~\cite{DBLP:conf/esop/GeorgesMOP17}, Sano
\etal~\cite{DBLP/conf/icfp/SanoGKBT2025}, and Ângelo
\etal~\cite{DBLP:journals/corr/abs-2404-05475}. Here we provide for concurrency,
recursive types, linear and unrestricted types, session types, and a programming
model encompassing a single (functional) programming language.


Fuses, by Sano \etal~\cite{DBLP/conf/icfp/SanoGKBT2025}, integrates contextual
modal types with session types in a manner different from ours. Their system
employs contextual modal types to separate a functional layer from a
session-typed process' layer, offering an alternative approach where linearity
is addressed only at the process level. Even if inspired in linear logic, Fuses,
as our system, offers no guarantees of deadlock freedom. Furthermore even if the
process layer can exchange recursive functional code in messages, it can't
splice the result into code with resorting to the functional level. Our
proposal, on the other hand offers a much more streamlined programming
experience with all the programming happening at the functional level. Fuses,
however, offers code analysis and pattern matching, topics that we decided to
leave to future work.


\section{Conclusion and future work}
\label{section:conclusion}
We show the merits of enhancing a session type system with staged
meta-programming, allowing protocols that exchange code. Such an extension
requires the integration of staged meta-programming into a linear type system
(on top of which session types are defined), accounting for the main technical
challenge. We also show our system can be implemented in a practical way, by
providing type checking rules, as well as correctness results. We further plan
on improving the system's expressivity, both on the way protocols are defined
and on how code is created and used. One could aim at two orthogonal extensions.
The first is polymorphism: both for session types
\cite{DBLP:journals/iandc/AlmeidaMTV22} and for staged meta-programming
\cite{DBLP:journals/pacmpl/JangGMP22}. The second is the incorporation of
context-free session types, allowing programming with the sequential composition
of types~\cite{DBLP:conf/esop/PocasCMV23,DBLP:conf/concur/SilvaMV23}. These
extensions would lead to more general and reusable code and protocols.




\paragraph{Acknowledgements}

\begin{sloppypar}
  We thank Chuta Sano for his insightful comments.
  This work was partly supported by JSPS Invitational Short-Term Fellowships for
  Research in Japan.
  It was further supported by the FCT through project Reliable and Expressive
  Concurrent Systems through Advanced Session Types ref.\ 2023.13752.PEX (\url{https://doi.org/10.54499/2023.13752.PEX}), and the LASIGE Research Unit,
  ref.\ UID/00408/2025 (\url{https://doi.org/10.54499/UID/00408/2025}),
  and by the LIACC Research Unit
  (\href{https://doi.org/10.54499/UIDB/00027/2020}{10.54499/UIDB/00027/2020} and
  \href{https://doi.org/10.54499/UIDP/00027/2020}{10.54499/UIDP/00027/2020}).
\end{sloppypar}


\bibliographystyle{splncs04}
\bibliography{biblio.bib} 

\newpage
\appendix
\section{Programming with multiple levels}
\label{sec:extra-examples}

Contextual types are indexed by \emph{levels}. We have been writing
\lstinline|Stream vdash Unit| as an abbreviation for 
\lstinline|Stream vdash1 Unit|, a level-1 contextual type.
One may distinguish two kinds of code fragments: those with box types (like
\lstinline|[Stream vdash Stream]|) stored in level-0 variables, and those with
contextual types (like \mbox{\lstinline|Stream vdash Stream|)} stored in level-1
or higher variables. The opening example of \cref{section:motivation}
demonstrates that level-1 contextual values can be duplicated and discarded.
It is worth noting that level-0 resources with box types can be unrestricted as well. The following variant of \lstinline|sendTwice|
exemplifies the unrestricted nature of box types, where the level-0
variable \lstinline|z| with a box type
\lstinline|[Stream vdash1 Stream]| is used twice.

\begin{lstlisting}
sendTwice' : Stream -> Unit
sendTwice' = let z = box (y. send 5 (select More y))
             let box u = z in
             let box v = z in
             in lambdax.close (select Done u[v[x]])
\end{lstlisting}

Code fragments are unrestricted resources in our type system because boxes are
designed to be independent of linear resources. Level-$n$ boxes abstract all
variables at levels lower than $n$, and this means that level-0 variables are
always abstracted. Our type system only allows level-0 variables to be linear.
Hence, boxes will not depend on linear resources, and it is safe to duplicate or
discard code fragments even when they are code fragments that compute linear
resources.

So far, all our examples showcase contextual types of up to level 1.
Here is an example featuring level 2 types, where code to send an arbitrary
number of integer values is sandwiched between code to send integer 100.
\begin{lstlisting}
sandwich : [Stream, (Stream vdash1 Stream) vdash2 Unit]
sandwich = box (x, u.
             let x1 = send 100 (select More x) in
             let x2 = u[x1] in
             let x3 = send 100 (select More x2) in
             close x3)
\end{lstlisting}
The communication template \lstinline|sandwich| is code parameterised on two arguments.
The first argument is a channel endpoint of type \lstinline|Stream|, which is a level 0 type.
The second is code parameterised on a channel endpoint that evaluates to another channel endpoint, both of type {\lstinline|Stream|}.
Then, the second argument is of level 1.
Hence, the type of the expression \lstinline|sandwich| is a level 2 box type.

Finally we splice in the code of \lstinline|sandwich| into function \lstinline|sandwichMe|, allowing both a channel endpoint and a supplied value, in the form of the box, to be spliced into the template code, filling the value to be sent in second.
The code
\lstinline|sandwichMe box(z. send 5 (select More (send 5 (select More z)))))|
results in sending four integer values (\lstinline{100}, \lstinline{5}, \lstinline{5} and \lstinline{100}) before closing the channel.

\begin{lstlisting}
sandwichMe : [Stream vdash1 Stream] -> [Stream vdash1 Unit]
sandwichMe y = let box u = y in
               let box v = sandwich in
               box (x. v[x, u])
\end{lstlisting}

\section{Proofs of the main results}
\label{sec:proofs}

\subsection{Basic results}

\begin{proof}[of \cref{lem:etactx}]
  By induction on the level $n$ of the contextual type $\tau$.
  
  Case $n = 0$. We have $\tau = (\contextualType{\emptyseq}0T)$ because
  $\tau$ is assumed to be well formed (that is, $\ctype \tau 0$ holds) and
  $\etactx{x}{\contextualType{\emptyseq}{0}{T}} =
  \ctxval{\emptyseq}{\appliedvar{x}{\epsilon}}$. By rule \ruletsvar,
  \isterm{\binding{x}{\tau}}{\appliedvar{x}{\epsilon}}{T}. By rule \ruletsctxt,
  \isterm{\binding{x}{\tau}}{\ctxval{\emptyseq}{\appliedvar{x}{\epsilon}}}{\tau}

  Case $n=k+1$. We have $\tau = (\contextualType{\overline\tau}{k+1}{T})$
  and
  $\etactx{x}{\tau} =
  \ctxval{\overline{y}}{\appliedvar{x}{\overline{\etactx{y}{\tau}}}}$.
  By induction we have
  $\overline{\isterm{\binding{y}{\tau}}{\etactx{y}{\tau}}{\tau}}$.
  By \ruletsvar we have
  $\isterm{\binding{x}{\tau},
    \overline{\binding{y}{\tau}}}{\appliedvar{x}{\etactx{y}{\tau}}}{T}$.
  Because $\tau$ is well formed, we have $\overline\tau < n$, and thus we can apply
  rule \ruletsctxt to obtain $ \isterm{\binding{x}{\tau}}{\ctxval{\overline{y}}{\appliedvar{x}{\etactx{y}{\tau}}}}{\tau}$.
\end{proof}

Let $\ctxgetun{\Gamma}$ denote the unrestricted part of $\Gamma$, that is, the
typing context containing exactly those bindings $\binding{x}{\ctxun{\tau}}$ in
$\Gamma$. We define $\ctxgetlin{\Gamma}$ in a corresponding manner.

\begin{lemma}[Basic properties of context split~\cite{DBLP:journals/iandc/Vasconcelos12}]
\label{lemm:props-split}
Let $\Gamma = \Gamma_1 \csplit \Gamma_2$.
\begin{enumerate}
\item $\ctxgetun{\Gamma} = \ctxgetun{\Gamma_1} = \ctxgetun{\Gamma_2}$.
\item If $\binding{x}{\ctxlin{\tau}} \in \Gamma$ then either $\binding{x}{\ctxlin{\tau}} \in \Gamma_1$ and $\binding{x}{\tau'} \not\in \Gamma_2$; or $\binding{x}{\ctxlin{\tau}} \in \Gamma_2$ and $\binding{x}{\tau'} \not\in \Gamma_1$, for any $\tau'$.
\item $\Gamma = \Gamma_2 \csplit \Gamma_2$.
\item If $\Gamma_1 = \Delta_1 \csplit \Delta_2$ then $\Delta = \Delta_2 \csplit \Gamma_2$ and $\Gamma = \Delta_1 \csplit \Delta$.
\end{enumerate}
\end{lemma}

\subsection{Preservation}

\begin{lemma}[Unrestricted weakening]
  \label{lem:weakening}
  \begin{mathpar}
    \inferrule[\ruleweak]
    {\isterm{\Gamma}{M}{T}}
    {\isterm{\Gamma,\binding x \ctxun\tau}{M}{T}}
  \end{mathpar}
\end{lemma}
\begin{proof}
  By rule induction on $\isterm \Gamma M T$.
\end{proof}

\begin{lemma}[Unrestricted strengthening]
  \label{lem:strengthening}
  If $\isterm{\Gamma,\binding x \tau}{M}{T}$ and $x\notin\fv{M}$, then $\un \tau$
  and $\isterm{\Gamma}{M}{T}$.
\end{lemma}
\begin{proof}
  By rule induction on $\isterm{\Gamma,\binding x \tau} M T$.
\end{proof}

\begin{lemma}[Context typing]
  \label{lem:typing-contexts}
  $\isterm{\Gamma}{\evaluationContextApply EM}{T}$ if and only if
  $\Gamma = \Gamma_1 \csplit \Gamma_2$ and $\isterm {\Gamma_1} M U$ and
  $\isterm{\Gamma_2, \binding{x}{}{U}}{\evaluationContextApply{E}{x}}{T}$.
\end{lemma}
\begin{proof}
By structural induction on $E$.
\end{proof}

\begin{lemma}[Substitution]
  \label{lem:substitution}
  Let $\isterm{\Gamma}{\sigma}{\tau}$ and
  $\Gamma \geq n$ where $n$ is the level of $\tau$ and
  suppose $\lin{\tau}$ or $\un{\Gamma}$.
  \begin{enumerate}
  \item If $\isterm{\Delta,\binding x {\tau}}{M}{T}$ and $\Gamma \csplit \Delta$ is well defined, then $\isterm{\Gamma \csplit \Delta}{\substitution{\sigma}{x}M}{T}$.
  \item If $\isterm{\Delta,\binding x {\tau}}{\rho}{\tau'}$ and $\Gamma \csplit \Delta$ is well defined, then $\isterm{\Gamma \csplit \Delta}{\substitution{\sigma}{x}\rho}{\tau'}$.
  \end{enumerate}
\end{lemma}
\begin{proof}

    We prove the two statements simultaneously by lexicographic induction on the
  level $n$ of $\tau$ and the size of the derivation of
  $\isterm{\Delta,\binding x {\tau}}{M}{T}$ or
  $\isterm{\Delta,\binding x {\tau}}{\rho}{\tau'}$ with case analysis on the
  last typing rule used. In this proof, we write $\Gamma \setminus x$ for the
  typing environment obtained by removing the binding of $x$ from $\Gamma$ (if any).

  We outline the most interesting cases.

  \noindent \textbf{Case} rule \ruletsvar: We have
  $M = \appliedvar{y}{\overline \rho}$ and
  $\Delta,\binding x {\tau} = \Gamma_1 \csplit \cdots \csplit \Gamma_m \csplit (\ctxun{\Delta_0}, \binding y {(\contextualType{\overline{\tau}}{n'}{T})})$ and
    $\isterm{\overline \Gamma}{\overline \rho}{\overline \tau}$
    for some $y, \overline \rho, \Delta_0, \overline \Gamma, \overline \tau, n'$.
  There are two possible sub-cases to consider:
  \begin{itemize}
  \item Subcase $x\neq y$:  We will show that
    $\isterm{\Gamma \csplit \Delta}{\appliedvar{y}{\substitution{\sigma}{x}\overline \rho}}{T}$ since
    $\substitution{\sigma}{x}{(\appliedvar{y}{\overline \rho})} = \appliedvar{y}{\substitution{\sigma}{x}\overline \rho}$.
    \begin{itemize}
    \item Subsubcase $\lin{\tau}$:  It must be the case that there exists $i$ such that $\Gamma_i = \Gamma'_i, x:\tau$ for some $\Gamma'_i$ and
      for $j \neq i$, $x \not \in \dom{\Gamma_j}$.
      It is easy to show that $\Gamma \csplit \Gamma'_j$ is well defined, because we assumed that $\Gamma \csplit \Delta$ is well defined.
      By the induction hypothesis, $\isterm{\Gamma \csplit \Gamma'_i}{\substitution{\sigma}{x}\rho_i}{\tau_i}$.
      For $j \neq i$, $\substitution{\sigma}{x}\rho_j = \rho_j$ as
      $x \not \in \dom{\Gamma_j}$ and thus $x$ is not a free variable of $\rho_j$.
      By \ruletsvar, we have
      $\isterm{(\Gamma_1 \csplit \cdots \csplit (\Gamma \csplit \Gamma'_i) \csplit \cdots \csplit \Gamma_m) \csplit
        (\ctxun{\Delta_0}, \binding y (\contextualType{\overline{\tau}}{n'}{T}))}{\appliedvar y {\substitution{\sigma}{x}\overline \rho}}{T}$.
      It is easy to show $(\Gamma_1 \csplit \cdots \csplit (\Gamma \csplit \Gamma'_j) \csplit \cdots \csplit \Gamma_m) \csplit
        \Gamma \csplit (\ctxun{\Delta_0}, \binding y (\contextualType{\overline{\tau}}{n'}{T})) = \Gamma \csplit (\Delta, \binding x \tau) \setminus x = \Delta$.
      
      \item Subsubcase $\un{\tau}$.  It must be the case that, for any
        $1 \leq i \leq m$, $\Gamma_i = \Gamma'_i, x:\tau$ for some
        $\Gamma'_i$.  We also have $\un{\Gamma}$ by assumption.
        Similarly to the above case,
        by the induction hypothesis, for any $i$, $\isterm{\Gamma \csplit \Gamma'_i}{\substitution{\sigma}{x}\rho_i}{\tau_i}$.
      By \ruletsvar, we have
      $\isterm{((\Gamma \csplit \Gamma'_1) \csplit \cdots \csplit (\Gamma \csplit \Gamma'_m) ) \csplit
        (\ctxun{\Delta_0}, \binding y (\contextualType{\overline{\tau}}{n'}{T}))}{\appliedvar y {\substitution{\sigma}{x}\overline \rho}}{T}$.
      It is easy to show $
((\Gamma \csplit \Gamma'_1) \csplit \cdots \csplit (\Gamma \csplit \Gamma'_m) ) \csplit
        (\ctxun{\Delta_0}, \binding y (\contextualType{\overline{\tau}}{n'}{T}))= 
\Gamma \csplit (\Gamma'_1 \csplit \cdots \csplit \Gamma'_m) \csplit
        (\ctxun{\Delta_0}, \binding y (\contextualType{\overline{\tau}}{n'}{T}))
       = \Gamma \csplit (\Delta, \binding x \tau) \setminus x = \Gamma \csplit \Delta$.

    \end{itemize}
  \item Subcase $x = y$:   We have
    $\tau = \contextualType{\overline{\tau}}{n'}{T}$ and
    $\sigma = \ctxval{\overline z}{N}$ for some $\overline z$ and $N$ and
    $\substitution{\sigma}{x}M = \substitution{\substitution{\sigma}{x}{\overline \rho}}{\overline z}{N}$.
    We will show that
    $\isterm{\Gamma \csplit \Delta}{\substitution{\substitution{\sigma}{x}{\overline \rho}}{\overline z}{N}}{T}$ .
    By \ruletsctxt, we have
    $\isterm{\Gamma, \overline{\binding z \tau}}{N}{T}$.  Note that $\ctxouter \Gamma n$ and
    $\ctxlocal {(\overline{\binding z \tau})}{n}$.  We have further case analysis on whether $n = 0$ or not
    and the muliplicity of $\tau$.
    \begin{itemize}
    \item Subsubcase $n = 0$.  In this case, $\overline z$ must be empty.  It suffices to show
      $\isterm{\Gamma \csplit \Delta}{N}{T}$ but it follows from Weakening.
      (It is easy to show $\un{\Delta}$ because $\overline \Gamma$ is also empty
      and $\Delta = \Delta_0$.)
    \item Subsubcase $n > 0$ and $\lin{\tau}$.  Similarly to the case for $x \neq y$ and $\lin{\tau}$,
      there exists $i$ such that
      $\Gamma_i  = \Gamma'_i, x : \tau$ and
      $\isterm{\Gamma \csplit \Gamma'_i}{\substitution{\sigma}{x}\rho_i}{\tau_i}$ and,
      for $j \neq i$, $\isterm{\Gamma'_j}{\substitution{\sigma}{x}\rho_j}{\tau_j}$.
      By the induction hypothesis (note that the levels of $\tau_j$ is strictly lower than $n$),
      $\isterm{\Gamma_1 \csplit \cdots \csplit (\Gamma \csplit \Gamma'_i) \csplit \cdots \csplit \Gamma_m}{\substitution{\substitution{\sigma}{x}{\overline \rho}}{\overline z}{N}}{T}$.
      It is easy to show that
      $\Delta =
        \Gamma_1 \csplit \cdots \csplit \Gamma'_i \csplit \cdots \csplit \Gamma_m \csplit \ctxun{\Delta_0}$.
      Weakening finishes the case.
    \item Subsubcase $n > 0$ and $\un{\tau}$ is similar.
    \end{itemize}
  \end{itemize}

\noindent \textbf{Case} rule \ruletsarrowuni.
Let that $M = \abstraction{y}{}{M_0}$ and $T = \functionType{T'}{\multun}{U'}$.
We then have that $\isterm{\Gamma}{\sigma}{\tau}$ and $\isterm{\Delta, \binding{x}{\tau}}{\abstraction{y}{}{M_0}}{\functionType{T'}{\multun}{U'}}$.
We want to show $\isterm{\Gamma \csplit \Delta}{\substitution{\sigma}{x} (\abstraction{y}{}{M_0})}{\functionType{T'}{\multun}{U'}}$.
\\\\
We have that $\substitution{\sigma}{x} (\abstraction{y}{}{M_0}) = \abstraction{y}{}{\substitution{\sigma}{x}M_0}$.
By \ruletsarrowuni, we have that $\un{\Delta}$ and $\un{\tau}$ and $\isterm{\Delta, \binding{x}{\tau}, \binding{y}{(\contextualType{\emptyseq}{0}{T'})}}{M_0}{U'}$.
Without loss of generality, we can assume $y \not \in \dom{\Gamma}$.
Now we have two subcases depending on the multiplicity of the type of $y$.
\begin{itemize}
\item Subcase $\lin{(\contextualType{\emptyseq}{0}{T'})}$: 
$\Gamma \csplit (\Delta, \binding{y}{(\contextualType{\emptyseq}{0}{T'})})$ is well defined.
By the induction hypothesis, $\isterm{\Gamma \csplit (\Delta, \binding{y}{(\contextualType{\emptyseq}{0}{T'})})}{\substitution{\sigma}{x}M_0}{U'}$,
which is equivalent to
$\isterm{(\Gamma \csplit \Delta), \binding{y}{(\contextualType{\emptyseq}{0}{T'}})}{\substitution{\sigma}{x}M_0}{U'}$.
\item Subcase $\un{(\contextualType{\emptyseq}{0}{T'})}$: 
  $(\Gamma, \binding{y}{(\contextualType{\emptyseq}{0}{T'})}) \csplit (\Delta, \binding{y}{(\contextualType{\emptyseq}{0}{T'})})$ is well defined and
  equal to  $(\Gamma \csplit \Delta), \binding{y}{(\contextualType{\emptyseq}{0}{T'})}$.
By Weakening and the induction hypothesis, $\isterm{(\Gamma \csplit \Delta), \binding{y}{(\contextualType{\emptyseq}{0}{T'})}}{\substitution{\sigma}{x}M_0}{U'}$.
\end{itemize}
Since $\un{\tau}$, we have $\un{\Gamma}$ by assumption.
Thus, $\un{(\Gamma \csplit \Delta)}$.
By \ruletsarrowuni, $\isterm{\Gamma \csplit \Delta}{\substitution{\sigma}{x} (\abstraction{y}{}{M_0})}{\functionType{T'}{\multun}{U'}}$.


\noindent \textbf{Case} rule \ruletsboxi.
We have $\Delta, \binding x {\tau} = \Delta_1, \Delta_2$ and $\un{\Delta_1}$ and $\un{\Delta_2}$ (hence $\un{\tau}$ and $\un{\Gamma}$) and $M = \cbox{\rho}$ and $T = \contextualModalType{(\contextualType{\overline{\tau'}}{n}{U})}$ and $\isterm{\Delta_1}{\rho}{(\contextualType{\overline{\tau'}}{n}{U})}$.
Since $\Gamma \csplit \Delta$ is well defined by assumption and
$\un{\Gamma}$ and $\un{\Delta}$, it must be the case that
$\Gamma = \Delta$ and $\Gamma \csplit \Delta = \Gamma$.
We want to show $\isterm{\Delta}{\substitution{\sigma}{x}{\cbox{\rho}}}{\contextualModalType{(\contextualType{\overline{\tau'}}{k}{U})}}$.

By Weakening,
$\isterm{\Delta, \binding x {\tau}}{\rho}{(\contextualType{\overline{\tau'}}{n}{U})}$.
By the induction hypothesis,
$\isterm{\Delta}{\substitution{\sigma}{x}{\rho}}
{(\contextualType{\overline{\tau'}}{n}{U})}$.
Rule \ruletsboxi finishes the case.



\noindent \textbf{Case} rule \ruletsboxe.
We have $M = \letbox{y}{M'}{N'}$ and
$\Delta, \binding x \tau = \Delta_1 \csplit \Delta_2$ and
$\isterm{\Delta_1}{M'}{\contextualModalType{\tau'}}$ and
$\isterm{\Delta_2, \binding y \tau'}{N'}{T}$
for some $y, M', N', \Delta_1, \Delta_2$.
We want to show
$\isterm{\Gamma \csplit \Delta}{\substitution{\sigma}{x}{(\letbox{y}{M'}{N'})}}{T}$.
We have that $\substitution \sigma x (\letbox y{M'}{N'}) = \letbox y{\substitution \sigma x{M'}}{\substitution \sigma x{N'}}$.
There are two possible sub-cases to consider:
\begin{itemize}
\item Subcase $\un{\tau}$:
  By \ruletsboxe,
  $\Delta_i = \Delta'_i, \binding{x}{\tau}$ (for $i = 1, 2$) for some
  $\Delta'_1, \Delta'_2$
  and $\isterm{\Delta'_1, \binding{x}{\tau}}{M'}{\contextualModalType{\tau'}}$ and $\isterm{\Delta'_2, \binding{x}{\tau^n}, \binding{y}{\tau'}}{N'}{T}$.
  By the induction hypothesis, $\isterm{\Gamma \csplit \Delta'_1}{\substitution \sigma x{M'}}{\contextualModalType{\tau'}}$ and $\isterm{\Gamma \csplit (\Delta'_2, \binding{y}{\tau'})}{\substitution \sigma x{N'}}{T}$.
  We have $\un{\Gamma}$ since $\un{\tau}$.
By \ruletsboxe and by \cref{lemm:props-split} , $\isterm{\Gamma \csplit (\Delta'_1 \csplit \Delta'_2)}{\letbox y{\substitution \sigma x{M'}}{\substitution \sigma x{N'}}}{T}$.  It is easy to see $\Delta = \Delta'_1 \csplit \Delta'_2$.

\item Subcase $\lin{\tau}$ and $x$ occurs free in $M'$:
  By \ruletsboxe,
  $\Delta_1 = \Delta'_1, \binding{x}{\tau}$ for some
  $\Delta'_1$ and
  $\isterm{\Delta'_1, \binding{x}{\tau}}{M'}{\contextualModalType{\tau'}}$ and $\isterm{\Delta_2, \binding{y}{\tau'}}{N'}{T}$.
By the induction hypothesis, $\isterm{\Gamma \csplit \Delta'_1}{\substitution \sigma x{M'}}{\contextualModalType{\tau'}}$.
Since $\isterm{\Delta_2, \binding{y}{\tau'}}{N'}{T}$ and $\binding{x}{\tau} \not\in \Delta_2, \binding{y}{\tau'}$, then $x$ does not occur free in $N'$ and hence, $\substitution \sigma x{N'} = N'$.
Therefore, $\isterm{\Delta_2, \binding{y}{\tau'}}{\substitution \sigma x{N'}}{T}$.
By \ruletsboxe and by \cref{lemm:props-split} , $\isterm{\Gamma \csplit (\Delta'_1 \csplit \Delta_2)}{\letbox y{\substitution \sigma x{M'}}{\substitution \sigma x{N'}}}{T}$.  It is easy to see $\Delta = \Delta'_1 \csplit \Delta_2$.
If instead $x$ occurs free in $N'$, the proof is similar.
\end{itemize}

\noindent \textbf{Case} rule \ruletsctxt:
We have $\rho = \ctxval{\overline y}M$ and
$\tau' = \contextualType{\overline{\tau'}}{k}{T'}$ for some $\overline y, M$ and
$(\Delta, \binding x \tau) \geq k$ and
$\isterm{\Delta, \binding x \tau, \overline{\binding y {\tau'}}}{M}{T'}$.
We want to show $\isterm{\Gamma \csplit \Delta}{\substitution \sigma x {\ctxval{\overline y}M}}{\contextualType{\overline{\tau'}}{k}{T'}}$.

Let $\overline{z : \tau''}$ be a subsequence of
$\overline{y : \tau'}$ with all $\un{\tau''_i}$.
Then, $(\Gamma, \overline{z : \tau''}) \csplit
(\Delta, \overline{y : \tau'}) = (\Gamma \csplit \Delta),
\overline{y : \tau'}$.
By weakening, we have $\isterm{\Gamma, \overline{z : \tau''}}{\sigma}{\tau}$.
By the induction hypothesis,
$\isterm{(\Gamma, \overline{z : \tau''}) \csplit (\Delta, \overline{\binding{y}{\tau'}})}{\substitution{\sigma}{x}M}{T'}$, which is equilavent to
$\isterm{(\Gamma \csplit \Delta), \overline{\binding{y}{\tau'}}}{\substitution{\sigma}{x}M}{T'}$.
Since $(\Delta, \binding x \tau) \geq k$, in particular $\tau \geq k$,
we have $\Gamma \geq k$.
By \ruletsctxt, $\isterm{\Gamma \csplit \Delta}{\ctxval{\overline{y}}{\substitution{\sigma}{x}M}}{\contextualType{\overline{\tau'}}{k}{T'}}$.
%
%
\end{proof}


\begin{proof}[Proof of \cref{thm:preservation-terms}]
By rule induction on $\reduction MN$.
The cases for \ruletrbeta and \ruletrletbox were discussed before.
The case for \ruletrsplit is similar to that of \ruletrbeta, using the fact that context split is commutative and associative (\cref{lemm:props-split}).
The case for \ruletrfix is straightforward.
The case for \ruletrctx follows from \cref{lem:typing-contexts}.
\end{proof}


\begin{lemma}[Unrestricted values are typed under unrestricted typing environments]
  \label{lem:unv-typed-untyenv}
  If $\isterm{\Gamma}v T$ and $\un T$, then $\un \Gamma$.
\end{lemma}
\begin{proof}
  By case analysis on the last typing rule used for $\isterm{\Gamma}v T$.
\end{proof}

\begin{lemma}[Value substitution]
  \label{lem:value-substitution}
  If $\isterm{\Gamma}{v}{U}$ and 
  $\isterm{\Delta,\binding x U}{M}{T}$ and $\Gamma \csplit \Delta$ is defined, then
  $\isterm{\Gamma \csplit \Delta}{\substitution{v}{x}M}{T}$.
\end{lemma}
\begin{proof}
  By \cref{lem:substitution,lem:unv-typed-untyenv}.
\end{proof}

\begin{lemma}[Soundness for structural congruence]
\label{lemma:soundness-sg}
Let $\processCong{P}{Q}$. Then $\isprocess{\Gamma}{P}$ if and only if $\isprocess{\Gamma}{Q}$.
\end{lemma}
\begin{proof}
By rule induction on $\isprocess{\Gamma}{P}$ and on $\isprocess{\Gamma}{Q}$.
\end{proof}

\subsection{Progress for the functional language}

\begin{proof}[Proof of \cref{lem:canonical-forms}]
  The proof is by case analysis on the last rule used in the derivation of
$\isterm{\Gamma^S}{v}{T}$.
\begin{enumerate}
\item The only rule that applies is \ruletsconst.
Then, we have that $v = c$ and $\Unit = \typeof{c}$.
According to \cref{figure:types-schemes-constants}, the only constant $c$ such that $\typeof{c} = \Unit$ is $\unit$.
Hence, $v = \unit$.
\item The only rule that applies is \ruletsvar.
We then have that $v = x[\overline{\sigma}]$ and $\isterm{(\ctxun\Delta , \overline{\Gamma} \csplit \binding{x}{(\contextualType{\overline{\tau}}{n}{T})})^S}{\appliedvar{x}{\overline\sigma}}{T}$.
Due to the restriction on contexts, we have that $\contextualType{\overline{\tau}}{n}{T} = \contextualType{\emptyseq}{0}{T}$, and hence $\overline{\sigma} = \emptyseq$.
Therefore, $v = x$, which is the abbreviated form of $\appliedvar{x}{\emptyseq}$.
\item The only rules that apply are:
\begin{itemize}
\item \ruletsconst.
Therefore, $v = \close$, $\wait$, $\send$, $\application{\send}{v}$, $\receive$, $\select{l}$, $\new$, $\fork$ or $\fix$.
\item \ruletsarrowlini or \ruletsarrowuni.
Therefore, $v = \abstraction{x}{M}$.
\end{itemize}
\item The only rule that applies is \ruletspairi.
Hence, $v = \pair u w$.
\item The only rule that applies is \ruletsboxi.
Hence, $v = \cbox{\sigma}$.
\end{enumerate}
\end{proof}


\begin{proof}[Proof of \cref{thm:progress}]
By rule induction on the hypothesis, using canonical forms (\cref{lem:canonical-forms}).
Cases for rules \ruletsconst, \ruletsvar, \ruletsarrowlini, \ruletsarrowuni, \ruletspairi and \ruletsboxi are trivial.

Case rule \ruletsarrowe.
Then $M = \application{M_1}{M_2}$ and $\isterm{{\Gamma_1}^S \csplit {\Gamma_2}^S}{\application{M_1}{M_2}}{T}$, with $\Gamma^S = {\Gamma_1}^S \csplit {\Gamma_2}^S$.
By \ruletsarrowe, $\isterm{{\Gamma_1}^S}{M_1}{\functionType{U}{m}{T}}$ and $\isterm{{\Gamma_2}^S}{M_2}{U}$.
By the induction hypothesis, we have that, for both $M_1$ and $M_2$, either 1., 2., or 3. apply.
There are several cases:
\begin{itemize}
\item both $M_1$ and $M_2$ are values.
Then, by \cref{lem:canonical-forms}, $M_1$ is $\abstraction{x}{M_1'}$, $\close$, $\wait$, $\send$, $\application{\send}{v}$, $\receive$, $\select{l}$, $\new$, $\fork$ or $\fix$.
If $M_1 = \abstraction{x}{M_1'}$ then by \ruletrbeta, $\reduction M N$, for some $N$.
If $M_1$ is $\close$, $\wait$, $\application{\send}{v}$, $\receive$ or $\select{l}$, then by \ruletsconst, $U$ is $\Close$, $\Wait$, $\modeType{\sendOp}{T}{S}$, $\modeType{\receiveOp}{T}{S}$ or $\choiceType{\internalChoiceOp}{l : S_l}{l \in L}$, respectively.
By \cref{lem:canonical-forms}, $M_2 = x$, and 3. applies.
If $M_1$ is $\send$, then $M$ is a value.
If $M_1$ is $\new$, then 3. applies.
If $M_1$ is $\fork$, then 3. applies
If $M_1$ is $\fix$, then by \ruletrfix, $\reduction M N$ for some $N$.
\item $M_1$ is a value and $\reduction{M_2 }{N_2}$, for some $N_2$.
Then, by \ruletrctx, $\reduction M N$ for some $N$.
\item $M_1$ is a value and $M_2$ fulfils 3.
Then 3. applies.
\item $\reduction {M_1}{N_1}$, for some $N_1$.
Then, by \ruletrctx, $\reduction M N$ for some $N$.
\item $M_1$ fulfils 3.
Then 3. applies.
\end{itemize}

Case rule \ruletspaire.
Then $M = \pairDestructor{x}{y}{M_1}{M_2}$ and $\isterm{\Gamma_1^S \csplit \Gamma_2^S}{M}{T}$.
By \ruletspaire, we have $\isterm{\Gamma_1^S}{M_1}{\productType{U_1}{U_2}}$ and $\isterm{\Delta^S,\binding{x}{(\contextualType \emptyseq 0 S_1)}, \binding{y}{(\contextualType \emptyseq 0 S_2)}}{M_2}{T}$.
By the induction hypothesis, for both $M_1$ and $M_2$, either 1., 2., or 3. apply.
There are several cases:
\begin{itemize}
\item $M_1$ is a value.
Then, by \cref{lem:canonical-forms}, $M_1 = \pair{u}{w}$ and $\reduction {M}{N}$ by \ruletrsplit.
\item $\reduction {M_1} {N_1}$, for some $N_1$.
Then, by \ruletrctx, $\reduction M N$, for some $N$.
\item $M_1$ fulfils 3.
Then we have that 3. applies.
\end{itemize}

Case rule \ruletsboxe.
Proof similar to the previous case.

Case rule \ruletsmatch.
Proof similar to the previous case.  
\end{proof}


\subsection{Absence of runtime errors for the process language}

\begin{lemma}[Canonical forms]
  \label{lem:canonical-forms}
  Let $\isterm{\Gamma^S}{v}{T}$.
  \begin{enumerate}
  \item If $T = \Unit$ then $v = \unit$.
  \item If $T = \Close, \Wait, \modeType{\sendOp}{T}{S}, 
    \modeType{\receiveOp}{T}{S},
    \choiceType{\internalChoiceOp}{l\colon S_l}{l \in L}$ or $\choiceType{\externalChoiceOp}{l\colon S_l}{l
      \in L}$, then $v = \appliedvar x \emptyseq$.
  \item If $T = \functionType{U}{\multOp}{V}$ then $v$ is $\abstraction{x}{M}$,
    $\close$, $\wait$, $\send$, $\application{\send}{v}$, $\receive$,
    $\select{l}$, $\new$, $\fork$ or $\fix$.
  \item If $T = \productType UV$ then $v = \pair uw$.
  \item If $T = \contextualModalType{\tau}$ then $v = \cbox \sigma$.
  \end{enumerate}
\end{lemma}
\begin{proof}
  By case analysis on the last rule used in the derivation of
  $\isterm{\Gamma^S}{v}{T}$.
\end{proof}

\begin{proof}[Proof for \cref{thm:progress}]
  By rule induction on the hypothesis, using canonical forms
  (\cref{lem:canonical-forms}).
\end{proof}

\begin{proof}[Proof for \cref{thm:absence-runtime-errors}]
Take 
$P \equiv \channelBinding{x_1}{y_1}{\cdots\channelBinding{x_n}{y_n}{(P_1 \PAR \cdots \PAR P_m)}}$.
Soundness for structural congruence (\cref{lemma:soundness-sg}) guarantees that the latter process is also typable under the empty context.
Now take $i,j,k$ such that the subject of $P_i$ is $x_k$, the subject of $P_j$ is $y_k$.
We know that there is a $\Gamma^S$ such that $\isprocess{\Gamma^S}{P_i \PAR P_j}$ since all entries in $\Gamma$ are introduced by rule \rulepfbind.
The same rule also ensures that $\Gamma^S$ contains entries of the form $\binding{x_k}{}{R}, \binding{y_k}{}{S}$ with $\dualRelation{R}{S}$.
Duality guarantees that $\channelBinding{x_k}{y_k}{(P_i \PAR P_j)}$ is a redex, hence $P$ is not a runtime error.
\end{proof}

\subsection{Proofs for algorithmic type checking (\cref{section:algorithmic-type-checking})}

\begin{lemma}[Properties of context difference]
\label{lemma:properties-context-difference}
Let $\ctxdiff{\Gamma}{x} = \Delta$.
\begin{enumerate}
\item $\Delta = \Gamma \setminus \{\binding{x}{\tau}\}$
\item $\ctxgetlin{\Gamma} = \ctxgetlin{\Delta}$
\item If $\binding{x}{\tau} \in \Gamma$, then $\ctxun\tau$ and $\binding{x}{\tau'} \not\in \Delta$, for some $\tau'$.
\end{enumerate}
\end{lemma}
\begin{proof}
The proof follows by case analysis.
\end{proof}

\begin{lemma}[Algorithmic monotonicity]
\label{lemma:algorithmic-monotonicity}
If $\typesynth{\Gamma}{M^\bullet}{T}{\Delta}$ then $\ctxgetun{\Gamma} = \ctxgetun{\Delta}$ and $\ctxgetlin{\Delta} \subseteq \ctxgetlin{\Gamma}$.
\end{lemma}
\begin{proof}
The proof follows by induction on the rules of $\typesynth{\Gamma}{M^\bullet}{T}{\Delta}$, using \cref{lemma:properties-context-difference}.
\end{proof}

\begin{lemma}[Algorithmic linear strengthening]
\label{lemma:algorithmic-linear-strengthening}
Assume $\lin{\tau}$.
If $\typesynth{\Gamma, \binding{x}{\tau}}{M^\bullet}{T}{\Delta, \binding{x}{\tau}}$ then $\typesynth{\Gamma}{M^\bullet}{T}{\Delta}$.
\end{lemma}
\begin{proof}
The proof follows by induction on the rules of $\typesynth{\Gamma}{M^\bullet}{T}{\Delta}$, using \cref{lemma:properties-context-difference} and \cref{lemma:algorithmic-monotonicity}.
\end{proof}

\begin{lemma}[Algorithmic weakening]
\label{lemma:algorithmic-weakening}
If $\typesynth{\Gamma}{M^\bullet}{T}{\Delta}$ then $\typesynth{\Gamma, \binding{x}{\tau}}{M^\bullet}{T}{\Delta, \binding{x}{\tau}}$.
\end{lemma}
\begin{proof}
The proof follows by induction on the rules of $\typesynth{\Gamma}{M^\bullet}{T}{\Delta}$, using \cref{lemma:properties-context-difference}.
\end{proof}

\begin{proof}[Proof of \cref{theorem:algorithmic-soundness}, Item 1]
The proof follows by induction on the rules of $\typesynth{\Gamma}{M^\bullet}{T}{\Delta}$, using \cref{lemma:properties-context-difference}, \cref{lemma:algorithmic-monotonicity} and \cref{lemma:algorithmic-linear-strengthening}.
The proof for rule \ruleatconst is trivial.
The proofs for rules \ruleatvarun, \ruleatvarlin, \ruleatarrowe, \ruleatpairi, \ruleatpaire and \ruleatmatch follow the same strategy as the proof for rule \ruleatboxe.
We detail the following cases:
\begin{itemize}
\item Rule \ruleatarrowlini.
Assume $\ctxun{(\ctxdiff{\Delta}{x})}$ and $\typesynth{\Gamma}{\absT{\multlin}{x}{T}{M^\bullet}}{\functionType{T}{\multlin}{U}}{\ctxdiff{\Delta}{x}}$.
By \cref{lemma:properties-context-difference}, we that $\ctxgetlin{\Delta} = \ctxgetlin{\ctxdiff{\Delta}{x}}$, hence $\ctxun{\Delta}$.
By rule \ruleatarrowlini, $\typesynth{\Gamma, \binding{x}{(\contextualType{\emptyseq}{0}{T})}}{M^\bullet}{U}{\Delta}$.
By the induction hypothesis, we have $\isterm{\Gamma,\binding{x}{(\contextualType{\emptyseq}{0}{T})}}{\erase{M^\bullet}}{U}$.
By rule \ruletsarrowlini, $\isterm{\Gamma}{\abstraction{x}{}{\erase{M^\bullet}}}{\functionType{T}{\multlin}{U}}$.
By erasure, $\isterm{\Gamma}{\erase{\absT{\multlin}{x}{T}{M^\bullet}}}{\functionType{T}{\multlin}{U}}$.

\item Rule \ruleatarrowuni.
Assume $\ctxun{\Gamma}$ and $\typesynth{\Gamma}{\absT{\multun}{x}{T}{M^\bullet}}{\functionType{T}{\multun}{U}}{\Gamma}$.
By rule \ruleatarrowuni, $\typesynth{\Gamma,\binding{x}{(\contextualType{\emptyseq}{0}{T})}}{M^\bullet}{U}{\Delta}$ and $\ctxdiff{\Delta}{x} = \Gamma$.
By \cref{lemma:properties-context-difference}, we that $\ctxgetlin{\Delta} = \ctxgetlin{\Gamma}$, hence $\ctxun{\Delta}$.
By the induction hypothesis, we have $\isterm{\ctxun\Gamma,\binding{x}{(\contextualType{\emptyseq}{0}{T})}}{\erase{M^\bullet}}{U}$.
By rule \ruletsarrowuni, $\isterm{\ctxun\Gamma}{\abstraction{x}{}{\erase{M^\bullet}}}{\functionType{T}{\multun}{U}}$.
By erasure, $\isterm{\Gamma}{\erase{\absT{\multun}{x}{T}{M^\bullet}}}{\functionType{T}{\multun}{U}}$.

\item Rule \ruleatboxi.
Assume $\ctxun{\Gamma}$ and $\typesynth{\Gamma}{\cbox{(\ctxval{\overline{\binding{x}{\tau}}}{M^\bullet})}}{\contextualModalType{(\contextualType{\overline{\tau}}{n}{T})}}{\Gamma}$.
By \ruleatboxi, $\typesynth{\Gamma}{\ctxval{\overline{\binding{x}{\tau}}}{M^\bullet}}{(\contextualType{\overline{\tau}}{n}{T})}{\Gamma}$.
The only rule that applies is \ruleatctxt, hence $\typesynth{\Gamma_1, \overline{\binding{x}{\tau}}}{M^\bullet}{T}{\Delta}$ with $\ctxdiv{\Gamma}{n}{\ctxouter{\Gamma_1}{n},\ctxlocal{\Gamma_2}{n}}$ and $\Gamma = \ctxdiff{\Delta}{\overline{x}}, \Gamma_2$.
It follows naturally that $\ctxun{(\ctxdiff{\Delta}{\overline{x}})}$ and $\ctxun{(\Gamma_2)}$.
Furthermore, by \cref{lemma:properties-context-difference}, we have that $\ctxgetlin{\Delta} = \ctxgetlin{\ctxdiff{\Delta}{x}}$, and thus $\ctxun{\Delta}$.
By the induction hypothesis, $\isterm{\ctxouter{\Gamma_1}{n}, \ctxlocal{\overline{\binding{x}{\tau}}}{n}}{\erase{M^\bullet}}{T}$.
By rule \ruletsctxt, $\isterm{\ctxouter{\Gamma_1}{n}}{\ctxval{\overline x}{\erase{M^\bullet}}}{(\contextualType{\overline \tau}{n}{T})}$.
By rule \ruletsboxi, $\isterm{\Gamma_1, \Gamma_2}{\cbox{(\ctxval{\overline x}{\erase{M^\bullet}})}}{\contextualModalType{(\contextualType{\overline \tau}{n}{T})}}$.
By erasure, $\isterm{\Gamma_1, \Gamma_2}{\erase{\cbox{(\ctxval{\overline{\binding{x}{\tau}}}{M^\bullet})}}}{\contextualModalType{(\contextualType{\overline \tau}{n}{T})}}$.

\item Rule \ruleatboxe.
Assume $\ctxun{(\ctxdiff{\Gamma_3}{x})}$ and $\typesynth{\Gamma_1}{\letbox{x}{M^\bullet}{N^\bullet}}{T}{\ctxdiff{\Gamma_3}{x}}$.
By \cref{lemma:properties-context-difference}, we have that $\ctxgetlin{\Gamma_3} = \ctxgetlin{\ctxdiff{\Gamma_3}{x}}$, and thus $\ctxun{(\Gamma_3)}$.
By \ruleatboxe, we have $\typesynth{\Gamma_1}{M^\bullet}{\contextualModalType{\tau}}{\Gamma_2}$ and $\typesynth{\Gamma_2, \binding x {\tau}}{N^\bullet}{T}{\Gamma_3}$.
By \cref{lemma:algorithmic-linear-strengthening}, we have that $\typesynth{\Gamma_1 \setminus \ctxgetlin{\Gamma_2}}{M^\bullet}{\contextualModalType{\tau}}{\Gamma_2 \setminus \ctxgetlin{\Gamma_2}}$.
By the induction hypothesis, $\isterm{\Gamma_1 \setminus \ctxgetlin{\Gamma_2}}{\erase{M^\bullet}}{\contextualModalType{\tau}}$ and $\isterm{\Gamma_2, \binding x {\tau}}{\erase{N^\bullet}}{T}$.
By \ruletsboxe, $\isterm{{\Gamma_1 \setminus \ctxgetlin{\Gamma_2}} \csplit \Gamma_2}{\letbox{x}{\erase{M^\bullet}}{\erase{N^\bullet}}}{T}$, and erasure, $\letbox{x}{\erase{M^\bullet}}{\erase{N^\bullet}} = \erase{\letbox{x}{M^\bullet}{N^\bullet}}$.
By \cref{lemma:algorithmic-monotonicity}, $\ctxgetun{\Gamma_1} = \ctxgetun{\Gamma_2}$ and $\ctxgetlin{\Gamma_2} \subseteq \ctxgetlin{\Gamma_1}$.
By \cref{lemm:props-split}, we have that ${\Gamma_1 \setminus \ctxgetlin{\Gamma_2}} \csplit \Gamma_2 = \Gamma_1$.
\end{itemize}
\end{proof}

\begin{proof}[Proof of \cref{theorem:algorithmic-completeness}, Item 2]
The proof follows by induction on the rules of $\isterm{\Gamma}{M}{T}$, using \cref{lemma:algorithmic-monotonicity} and \cref{lemma:algorithmic-weakening}.
The proof for rule \ruletsconst, \ruletsarrowlini and \ruletsarrowuni are trivial.
The proofs for rules \ruletsvar, \ruletsarrowe, \ruletspairi, \ruletspaire and \ruletsmatch follow the same strategy as the proof for rule \ruletsboxe.
For rule \ruletsvar, we must weaken the judgments obtained from induction hypothesis from left to right.
We detail the following cases:
\begin{itemize}
\item Rule \ruleatboxi.
Assume $\isterm{\ctxun\Gamma, \ctxun{\Gamma'}} {\cbox{(\ctxval{\overline{x}}{M})}}{\contextualModalType{(\contextualType{\overline{\tau}}{n}{T})}}$.
By \ruletsboxi and \ruletsctxt, we have that $\isterm{\ctxouter{\Gamma}{n},\ctxlocal{\overline{\binding{x}{\tau}}}{n}}{M}{T}$.
By the induction hypothesis, there is a $M^\bullet$ such that $\erase{M^\bullet} = M$, and $\typesynth{\ctxouter{\Gamma}{n},\ctxlocal{\overline{\binding{x}{\tau}}}{n}}{M^\bullet}{T}{\Delta}$ and $\ctxun{\Delta}$.
Lets split the unused context $\Gamma'$ in the following way: $\ctxdiv{\Gamma'}{n}{\ctxouter{(\Gamma'_1)}{n},\ctxlocal{(\Gamma'_2)}{n}}$.
Then we can apply \ruleatctxt followed by \ruleatboxi, to reach $\typesynth{\ctxouter{\Gamma}{n},\ctxlocal{\Gamma_2'}{n}}{\cbox{(\ctxval{\overline{\binding{x}{\tau}}}{M^\bullet})}}{\contextualModalType{(\contextualType{\overline{\tau}}{n}{T})}}{\ctxdiff{\Delta}{\overline{x}}, \ctxlocal{(\Gamma'_2)}{n}}$.
By \cref{lemma:algorithmic-weakening}, we weaken the algorithmic typing judgment with $\ctxouter{(\Gamma'_1)}{n}$.
Thus, we arrive at $\typesynth{\Gamma,\Gamma'}{\cbox{(\ctxval{\overline{x}}{M^\bullet})}}{\contextualModalType{(\contextualType{\overline{\tau}}{n}{T})}}{(\ctxdiff{\Delta}{\overline{x}}), \Gamma'}$.
It's easy to check that $\ctxun{((\ctxdiff{\Delta}{\overline{x}}), \Gamma')}$.

\item Rule \ruletsboxe.
Assume $\isterm{{\Gamma} \csplit \Delta}{\letbox{x}{M}{N}}{T}$.
Induction on the premises gives us $\typesynth{\Gamma} {M^\bullet}{\contextualModalType{\tau}}{\Gamma'}$ with $\ctxun{(\Gamma')}$ and $\erase{M^\bullet} = M$, and $\typesynth{\Delta,\binding x {\tau}} {N^\bullet}{T}{\Delta'}$ with $\ctxun{(\Delta')}$ and $\erase{N^\bullet} = N$.
Since $\Gamma \cup \ctxgetlin{\Delta} = \Gamma \csplit \Delta$, we weaken (\cref{lemma:algorithmic-weakening}) $\typesynth{\Gamma} {M^\bullet}{\contextualModalType{\tau}}{\Gamma'}$, resulting in $\typesynth{\Gamma \csplit \Delta} {M^\bullet}{\contextualModalType{\tau}}{\Gamma' \cup \ctxgetlin{\Delta}}$.
We have to show that $\Gamma' \cup \ctxgetlin{\Delta} = \Delta$.
Since $\ctxun{(\Gamma')}$, then $\ctxgetlin{\Gamma' \cup \ctxgetlin{\Delta}} = \ctxgetlin{\Delta}$.
By \cref{lemma:algorithmic-monotonicity}, we have that $\ctxgetun{\Gamma} = \ctxgetun{\Gamma'}$.
Therefore, $\ctxgetun{\Gamma' \cup \ctxgetlin{\Delta}} = \ctxgetun{\Gamma'} = \ctxgetun{\Gamma} = \ctxgetun{\Gamma \csplit \Delta} = \ctxgetun{\Delta}$.
Therefore, $\Gamma' \cup \ctxgetlin{\Delta} = \Delta$.
By \ruleatboxe, we have $\typesynth{\Gamma \csplit \Delta}{\letbox{x}{M^\bullet}{N^\bullet}}{T}{\ctxdiff{\Delta'}{x}}$.
Since $\ctxun{(\Delta')}$, by \cref{lemma:properties-context-difference}, $\ctxun{(\ctxdiff{\Delta'}{x})}$.
\end{itemize}
\end{proof}


\end{document}